\documentclass{aa}  
\usepackage{times} 
\usepackage[dvips,xdvi]{graphics}
\def\AV{\mbox{${A_{\rm V}}$}}
\def\HA{\mbox{H$\alpha$}}
\def\HB{\mbox{H$\beta$}}
\def\HG{\mbox{H$\gamma$}}
\def\HD{\mbox{H$\delta$}}
\def\BG{\mbox{Br$\gamma$}}
\def\mic{\mbox{$\mu$m}}

\def\LSUN{\mbox{$L_\odot$}}
\def\LFIR{\mbox{$L_{\rm FIR}$}}

\def\ZSUN{\mbox{$Z_\odot$}}
\def\L{\mbox{$\lambda$}}
\def\LL{\mbox{$\lambda\lambda$}}
\def\Ne{\mbox{$n_e$}}
\def\NH{\mbox{$n_{\rm H}$}}

\def\Te{\mbox{$T_e$}}

\begin{document}

\title{
Metal abundances and excitation of extranuclear clouds in the Circinus
galaxy
\thanks{Based on observations collected 
at the European Southern Observatory, La Silla, Chile}}
\subtitle{ 
A new method for deriving abundances of AGN narrow line clouds}

\author{E. Oliva\inst{1}, 
A. Marconi\inst{1}, 
and A.F.M Moorwood\inst{2} 
}

\institute{
Osservatorio Astrofisico di Arcetri, Largo E. Fermi 5, 
I--50125 Firenze, Italy
\and
European Southern Observatory, Karl Schwarzschild Str. 2, D-85748
Garching bei M\"unchen, Federal Republic of Germany 
}

\offprints{E. Oliva}

\date{Received 23 July 1998 }

\thesaurus{03?(                  
               11.01.1;          
               11.01.2;          
               11.09.1 Circinus; 
               11.19.1;          
               11.19.3           
             )}
\titlerunning{Metal abundances of AGN excited extranuclear clouds}
\authorrunning{Oliva, Marconi \& Moorwood }
\maketitle

\begin{abstract}
Spectra of extranuclear clouds within the ionization cone of Circinus 
galaxy are presented and
modelled assuming photoionization by the Seyfert 2 nucleus
but no preconceived assumption concerning
the spectral shape of the ionizing AGN radiation or
the gas density distribution.

The most important result is that, regardless of the assumed 
AGN spectral shape or density distribution, the metal abundances 
are remarkably well ($\pm$0.2 dex) constrained by the data.
Oxygen and neon are found to be a factor of 5 below solar,
while nitrogen is solar
(cf. Table~\ref{tab_Z}). The large N/O overabundance is in agreement with the
chemical enrichment expected from the old (circum)nuclear starburst.
Similar studies of extranuclear clouds in other objects could therefore
provide a powerful tool to determine metallicities and trace past
starburst activity in AGNs.

Other implications such as the role of shock excitation,
which is effectively excluded, and
the intrinsic shape of the AGN ionizing continuum,
which is poorly constrained by the data, 
are also discussed.

\end{abstract}

\section{Introduction}

The Circinus galaxy (A1409-65) is a nearby ($\simeq$4 Mpc) gas rich spiral
lying close to the galactic plane in a region of relatively low 
(\AV$\simeq$1.5 mag) interstellar extinction (Freeman et al. \cite{freeman}).
Several observed characteristics indicate that this galaxy hosts the nearest
Seyfert 2 nucleus known.
These include optical images showing a spectacular
[OIII] cone (Marconi et al. 1994 hereafter \cite{M94}); 
optical/IR spectra rich 
in prominent and narrow coronal lines (Oliva et al. 1994,
hereafter \cite{O94},
Moorwood et al. 1996, hereafter \cite{M96}); 
X-ray spectra displaying a very
prominent Fe-K fluorescent line (Matt et al. \cite{matt96}) and optical
spectropolarimetric data which reveal relatively
broad H$\alpha$ emission in polarized light (Oliva et al. \cite{oliva98}).
Complementary to these is observational evidence that this galaxy 
has recently experienced a powerful nuclear starburst which is now
traced by the near IR emission of red supergiants (Oliva et al. 
\cite{oliva95}, Maiolino et al. \cite{maiolino}) 
and which may have propagated outwards
 igniting the bright ring of O stars and
HII regions visible in the \HA\  image (\cite{M94}). 
Such a starburst could have been triggered by gas moving toward
the nuclear region and eventually falling onto the accretion
disk around the black hole powering the AGN.

A debated issue is whether nuclear starbursts are common
features of AGNs and if they are more common in type 2 
than in type 1 Seyferts, as suggested by e.g. 10$\mu$m observations
(Maiolino et al. \cite{maiolino95}) and studies of the stellar
mass to light ratios (\cite{O94}).
Since starbursts are predicted and observed to deeply modify the
chemical abundances of the host galaxy
(e.g. Matteucci \& Padovani \cite{matteucci93}), such an effect 
should also be evident in this and other Seyferts. 
However, to the best
of our knowledge, no reliable measurement of metallicity for 
the narrow line region clouds of Seyfert 2's
exists in the literature. In particular, 
although has been since long known that the large [NII]/H$\alpha$ ratio
typical of Seyferts cannot be easily explained using simple models 
with normal
nitrogen abundances (e.g. 
Osterbrock \cite{osterbrock89},
Komossa \& Schulz \cite{komossa97}),
the question of whether its absolute (N/H) or relative (e.g. N/O)
abundance is truly different than solar 
is still open.
Finding a reliable method to derive metallicities and, therefore, to
trace and put constraints on past starburst activity is the main aim 
of this paper.

\begin{table}
\caption{Observed and dereddened line fluxes}
\label{tab_obs1}
\vskip-9pt
\def\SKIP#1{\noalign{\vskip#1pt}}
\def\UNC{\rlap{:}}
\def\UNO{\ \rlap{$^{(1)}$}}
\def\DUE{\ \rlap{$^{(2)}$}}
\def\TRE{\rlap{$^{(3)}$}}
\begin{flushleft}
\begin{tabular}{lrrcrrcc}
\hline
\hline\SKIP{1}
    & \multicolumn{3}{c}{Nucleus}   &  \multicolumn{3}{c}{KnC}   \\
    & \multicolumn{3}{c}{(4.6"x2")} & \multicolumn{3}{c}{(4.1"x2")}   \\
    & \multicolumn{3}{c}{\AV=4.5}   & 
          \multicolumn{3}{c}{\AV=1.9}   \\
\SKIP{3}
  & F\UNO & I\DUE & & F\UNO & I\DUE & &
  ${\rm A_\lambda /A_V}$\TRE \\
\SKIP{1}
\hline
\SKIP{3}
  ${\rm   [OII]   }\,\lambda\lambda$3727
 &     77 &    270    
&
 &     78 &    133
&
 &1.49   \\
  ${\rm   [NeIII] }\,\lambda$3869
 &     45 &    136   
&
 &     41 &     66
&
 &1.45   \\
  ${\rm   [SII]   }\,\lambda\lambda$4073
 &  &
&
 &   4\UNC &    7\UNC
&
 &1.40   \\
  ${\rm \HD       }\,\lambda$4102
 & & 
&
 &     17 &     24
&
 &1.40   \\
  ${\rm \HG       }\,\lambda$4340
 & & 
&
 &     35 &     46
&
 &1.34   \\
  ${\rm   [OIII]  }\,\lambda$4363
 & & 
&
 &     16 &     21
&
 &1.34   \\
  ${\rm HeII      }\,\lambda$4686
 &     32 &     41
&
 &     54 &     60
&
 &1.24   \\
  ${\rm   [ArIV]  }\,\lambda$4711
 & & 
&
 &   9 &  10
&
 &1.23   \\
  ${\rm   [ArIV]  }\,\lambda$4740
 & & 
&
 &   9 &     10
&
 &1.22   \\
  ${\rm \HB       }\,\lambda$4861
 &    100 &    100
&
 &    100 &    100
&
 &1.18   \\
  ${\rm   [OIII]  }\,\lambda$4959
 &    365 &    320  
&
 &    317 &    300
&
 &1.15   \\
  ${\rm   [OIII]  }\,\lambda$5007
 &   1245 &   1025   
&
 &   1048 &    965
&
 &1.14   \\
   ${\rm   [FeVI]  }\,\lambda$5146
 & &
&
 & $<$8 & $<$7
&
 &1.09   \\
  ${\rm   [NI]    }\,\lambda\lambda$5199
 &     28 &     18
&
 &   9 &   7
&
 &1.08   \\
  ${\rm   [FeXIV] }\,\lambda$5303
 & $<$17 & $<$10
&
 & $<$7 & $<$5
&
 &1.05   \\
  ${\rm HeII      }\,\lambda$5411
 & & 
&
 &   5.9 &   4.4
&
 &1.02   \\
  ${\rm   [FeVII] }\,\lambda$5721
 &     23 &     9 
&
 &   8.9 &   6.0
&
 &0.95   \\
  ${\rm   [NII]   }\,\lambda$5755
 & & 
&
 &   7.2 &   4.7
&
 &0.95   \\
  ${\rm HeI       }\,\lambda$5876
 &     35\UNC &     12\UNC
&
 &     15 &   9.7
&
 &0.92   \\
  ${\rm   [FeVII] }\,\lambda$6087
 &     36 &     11 
&
 &     16 &   9.7
&
 &0.89   \\
  ${\rm   [OI]    }\,\lambda$6300
 &    165 &     42
&
 &     46 &     26
&
 &0.85   \\
  ${\rm   [SIII]  }\,\lambda$6312
 &     20\UNC &   5\UNC
&
 &   11 &   6
&
 &0.85   \\
  ${\rm   [OI]    }\,\lambda$6364
 &     56 &     14
&
 &     16 &   8.6
&
 &0.84   \\
  ${\rm   [FeX]   }\,\lambda$6374
 &     55 &     14
&
 & $<$4
 & $<$2
&
 &0.84   \\
  ${\rm   [ArV]   }\,\lambda$6435
 & & 
&
 &   5.4 &   2.9
&
 &0.83   \\
  ${\rm   [NII]   }\,\lambda$6548
 &    535 &    116
&
 &    154 &     81
&
 &0.81   \\
  ${\rm   \HA     }\,\lambda$6563
 &   1390 &    298
&
 &    565 &    295
&
 &0.81   \\
  ${\rm   [NII]   }\,\lambda$6583
 &   1620 &    343
&
 &    457 &    237
&
 &0.81   \\
  ${\rm HeI       }\,\lambda$6678
 &       &     
&
 &   5.2 &   2.6  
&
 &0.80   \\
  ${\rm   [SII]   }\,\lambda$6716
 &    490 &     96  
&
 &    128 &     64
&
 &0.79   \\
  ${\rm   [SII]   }\,\lambda$6731
 &    496 &     96
&
 &    113 &     57
&
 &0.79   \\
  ${\rm   [ArV]   }\,\lambda$7006
 &   10\UNC &  2\UNC
&
 &     14 &   6.4
&
 &0.75   \\
  ${\rm   [ArIII] }\,\lambda$7136
 &    148 &     22
&
 &     61 &     27
&
 &0.73   \\
  ${\rm   [CaII]  }\,\lambda$7291
 & $<$20 & $<$3
&
 & $<$4 & $<$2
&
 &0.70   \\
  ${\rm   [OII]   }\,\lambda\lambda$7319
 &     45 &   6   
&
 &     12 &   5.2
&
 &0.70   \\
  ${\rm   [OII]   }\,\lambda\lambda$7330
 &     40 &   5   
&
 &     10 &   4.5
&
 &0.70   \\
  ${\rm   [ArIII] }\,\lambda$7751
 &     38 &   4  
&
 &     12 &   4.8
&
 &0.64   \\
  ${\rm   [FeXI]  }\,\lambda$7892
 &    112 &     11
&
 & $<$9
 & $<$3
&
 &0.62   \\
  ${\rm   [FeII]  }\,\lambda$8617
 & $<$40
 & $<$3
&
 & $<$13
 & $<$4
&
 &0.52   \\
  ${\rm   [SIII]  }\,\lambda$9069
 &    980 &     53
&
 &    220 &     65
&
 &0.48   \\
  ${\rm   [SIII]  }\,\lambda$9531
 &   2900 &    134
&
 &    570 &    155 
&
 &0.44   \\
  ${\rm   [CI]    }\,\lambda$9850
 &     70 &   2.9
&
 & $<$33
 & $<$9
&
 &0.42   \\
  ${\rm   [SVIII] }\,\lambda$9913
 &    100 &   4.1
&
 & & 
&
 &0.41   \\
  ${\rm Pa7       }\,\lambda$10049
 &     90\UNC &   4\UNC
&
 & & 
&
 &0.40   \\
  ${\rm   [FeII]  }\,\lambda$16435
 &    470 &   7.1
&
 &     30\UNC &   5\UNC  
&
 &0.17   \\
  ${\rm   [SiVI]  }\,\lambda$19629
 &    640 &   8
&
 &    $<$90 &  $<$15
&
 &0.12   \\
  ${\rm H_2       }\,\lambda$21213
 &    400 &   4.7
&
 &     50\UNC &   8\UNC
&
 &0.11   \\
  ${\rm \BG       }\,\lambda$21655
 &    230 &   2.7
&
 &     18\UNC &   3\UNC
&
 &0.11   \\
\SKIP{2}
 \HB\ flux$^a$
 &    10 &  1350
&
 &     5 &    36 
& 
 & \\
\SKIP{1}
\hline
\end{tabular}
\def\NOTA#1#2{\hbox{\vtop{\hbox{\hsize=0.030\hsize\vtop{\centerline{#1}}}}
      \vtop{\hbox{\hsize=0.97\hsize\vtop{#2}}}}}
\vskip1pt
\NOTA{ $^{(1)}$ }{ Observed line flux, relative to \HB=100}
\NOTA{ $^{(2)}$ }{ Dereddened flux (\HB=100), a colon denotes uncertain 
values. Blank entries are undetected lines with non-significant upper limits
}
\NOTA{ $^{(3)}$ }{ From Savage \& Mathis 
(\cite{savage79}) and Mathis (\cite{mathis90})}
\NOTA{ $^a$}{ Units of $10^{-15}$ erg cm$^{-2}$ s$^{-1}$}
\end{flushleft}
\end{table}

\begin{figure}
\centerline{\resizebox{\hsize}{!}{\includegraphics{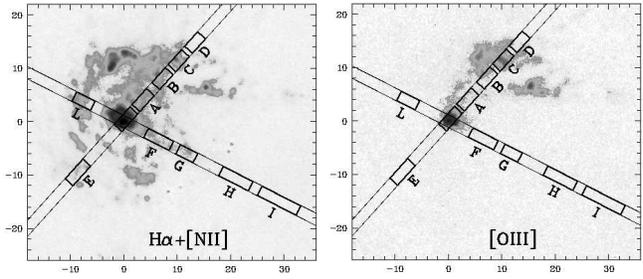}}}
\caption{ 
Slit positions overlaid on grey-scale line images. The positions
of the various knots are marked.
}
\label{showslit}
\end{figure}

\begin{figure}
\centerline{\resizebox{8.8cm}{!}{\includegraphics{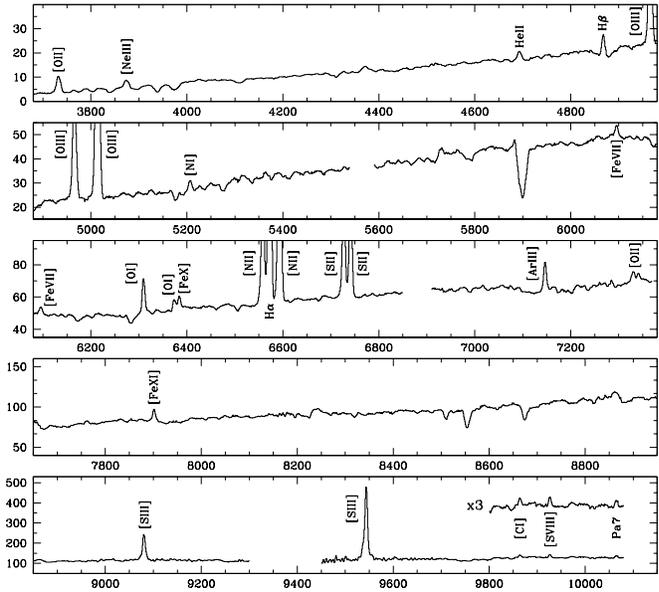}}}
\caption{ 
Spectrum of the nucleus, 
i.e. the central region 
at PA=318$^\circ$ (cf. Figs. 1, 4).
Fluxes are in units of 10$^{-16}$ erg cm$^{-2}$ s$^{-1}$ \AA$^{-1}$.
}
\label{spec_nuc}
\end{figure}

\begin{figure}
\centerline{\resizebox{8.8cm}{!}{\includegraphics{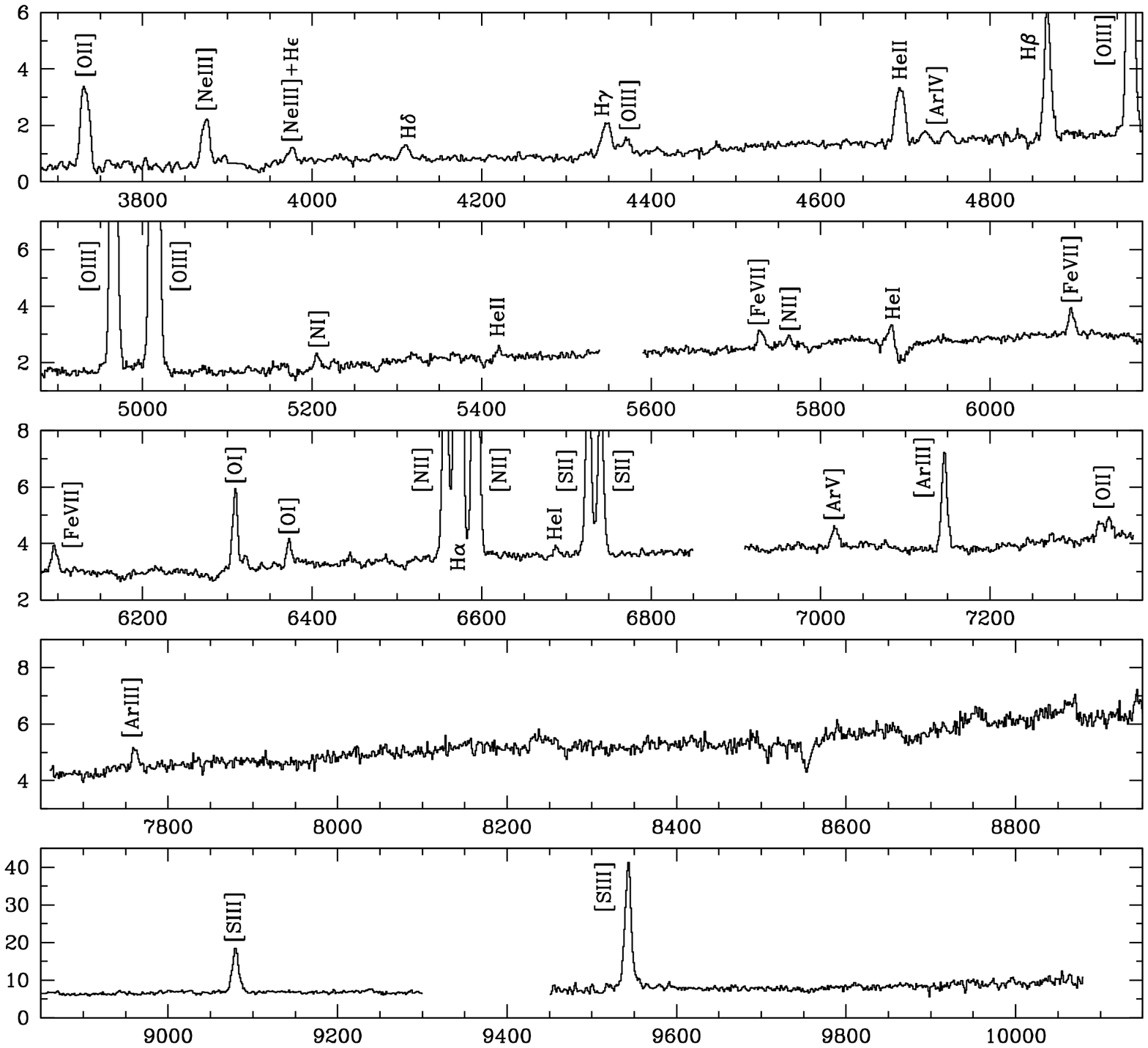}}}
\caption{
Spectrum of Knot C,
i.e. a  region 15.5\arcsec from
the nucleus at PA=318$^\circ$ (cf. Figs.~1, 4).
Flux units are as in Fig.~\ref{spec_nuc}
}
\label{spec_knc}
\end{figure}
We chose the Circinus galaxy as a benchmark
because its emission line spectrum is characterized by remarkably narrow 
($\la$150 km/s, \cite{O94}) emission lines which
are particularly easy to measure and which indicate relatively low
dynamical activity. This last aspect may be used to put tight constrains on
the possible contribution of shock excitation which may complicate 
the modelling of the observed spectrum and the determination of metallicities.

This paper presents new optical and infrared spectroscopic data
and is structured as follows.
Observations and data reduction are described in Sect. 2 and the results
are analyzed in Sect. 3.
In Sect. 4 we constrain the excitation conditions of the gas and
model the observed spectra in terms of photoionization from the AGN.
The derived chemical abundances are discussed in Sect. 5 and
in Sect. 6 we draw our conclusions.

\section {Observations and Data Reduction}

Long slit optical spectra were  collected 
at the ESO NTT telescope in March 1995 using a dichroic beam
splitter feeding both blue (1024$^2$ Tek-CCD with 0.37\arcsec/pix)  
and red (2048$^2$ Tek-CCD with 0.27\arcsec/pix) arms of EMMI.
Simultaneous blue 
(3700--5000 with $\simeq$1.7 \AA/pix) and red (5000--10000 with
$\simeq$1.2 \AA/pix) 
spectra were obtained through a 2\arcsec slit centered on the optical
nucleus at two position angles (cf. Fig.~\ref{showslit}).
The first was at PA=318$^\circ$, roughly aligned with the [OIII]
cone axis and including the brightest [OIII] emitting knots,
while the second was at PA=243$^\circ$ and along the low excitation
rim visible in the [SII] and
``true colour'' images shown in Fig.~5 and Fig.~10 of \cite{M94},
\footnote{These images are available at 
http://\-www.arcetri.astro.it/$\sim$oliva}.
Several exposures were averaged and
the total integration times of the spectra at PA=318$^\circ$ 
were 100, 75, 25 minutes over the 3700--5000, 5000--7300 and 7700--10080 \AA\
wavelength ranges, respectively. 
The spectra at PA=243$^\circ$ covered 3700-5000 and 5000--7300 \AA\
with a total integration time of 50 minutes per wavelength interval.
Data were flux calibrated using observations of LTT3218
and reduced with MIDAS using standard procedures.

Infrared spectra were collected in March 1995, also
at the ESO NTT using the spectrometer IRSPEC equipped with a 62x58
SBRC InSb array whose pixel size was 2.2\arcsec along the slit
and $\simeq$5 \AA\  along the dispersion direction.
Spectra of [FeII]1.644 \mic, H$_2$ 2.121 \mic, \BG\ and
 [SiVI]1.962 \mic\  were
obtained using a 4.4\arcsec slit at PA=318$^\circ$.
Each long-slit spectrum consisted of 4 ABBA cycles (A=source, B='sky',
i.e. a region 300\arcsec E) with 2x60 sec integrations per position.
The data were reduced using the IRSPEC context of MIDAS which was developed
by one of us (EO) and which also allows the accurate subtraction of 
time variable OH sky lines. The same airglow emission was used as
wavelength reference (Oliva \& Origlia \cite{oliva92}) and the spectra were
corrected for instrumental and atmospheric transmission using spectra
of featureless early O stars and flux calibrated using measurements
of photometric standard stars.

\begin{table*}
\caption{Fluxes of significant lines
\label{tab_obs2}
in the various knots$^{(1)}$ }
\def\SKIP#1{\noalign{\vskip#1pt}}
\def\S{\ \ \ \ \ }
\def\UNC{\rlap{:}}
\begin{flushleft}
\begin{tabular}{lrrrrrrrrrrr}
\hline\hline
 & \multicolumn{1}{c}{Nucleus} & \multicolumn{1}{c}{KnA} 
     & \multicolumn{1}{c}{KnB} & \multicolumn{1}{c}{KnC} 
     & \multicolumn{1}{c}{KnD} & \multicolumn{1}{c}{KnE} 
     & \multicolumn{1}{c}{KnF$^a$}  & \multicolumn{1}{c}{KnG$^a$} 
     & \multicolumn{1}{c}{KnH$^a$} & \multicolumn{1}{c}{KnI$^a$}
     & \multicolumn{1}{c}{KnL$^a$} \\
 & \multicolumn{1}{c}{4.6"x2"} & \multicolumn{1}{c}{4.6"x2"} 
     & \multicolumn{1}{c}{4.1"x2"} & \multicolumn{1}{c}{4.1"x2"} 
     & \multicolumn{1}{c}{4.1"x2"} & \multicolumn{1}{c}{5.4"x2"} 
     & \multicolumn{1}{c}{5.7"x2"}  & \multicolumn{1}{c}{4.1"x2"} 
     & \multicolumn{1}{c}{6.5"x2"} & \multicolumn{1}{c}{8.4"x2"}
     & \multicolumn{1}{c}{4.3"x2"} \\
 & \multicolumn{1}{c}{\llap{A}$_{\rm V}$=4.5}
& \multicolumn{1}{c}{\llap{A}$_{\rm V}$=4.4}
& \multicolumn{1}{c}{\llap{A}$_{\rm V}$=3.8}
& \multicolumn{1}{c}{\llap{A}$_{\rm V}$=1.9}
& \multicolumn{1}{c}{\llap{A}$_{\rm V}$=2.0}
& \multicolumn{1}{c}{\llap{A}$_{\rm V}$=5.0}
& \multicolumn{1}{c}{\llap{A}$_{\rm V}\!\simeq\!4$}
& \multicolumn{1}{c}{\llap{A}$_{\rm V}\!\simeq\!4$}
& \multicolumn{1}{c}{\llap{A}$_{\rm V}\!\simeq\!1.5$}
& \multicolumn{1}{c}{\llap{A}$_{\rm V}\!\simeq\!1.5$}
& \multicolumn{1}{c}{\llap{A}$_{\rm V}\!\simeq\!6$}
\\
\hline
\SKIP{1}
  ${\rm   [OII]   }\,\lambda\lambda$3727
 &    270\S
 &    740\rlap{$^b$}\S
 &    312\S
 &    133\S
 &    112\S
 & $<$200\S
 &    650\rlap{$^b$}\S
 &    400\S
 &    410\S
 &    380\S
 &    290\UNC\S
\\
  ${\rm HeII      }\,\lambda$4686
 &     41\S
 &     40\UNC\S
 &     30\UNC\S
 &     60\S
 &     53\S
 & $<$25\S
 &  
 & 
 & 
 & 
 & 
\\
  ${\rm   [OIII]  }\,\lambda$5007
 &   1025\S
 &    900\S
 &    690\S
 &    965\S
 &    815\S
 &     20\UNC\S
 &    400\S
 &    230\S
 &    260\S
 &    250\S
 &   
\\
  ${\rm   [FeVII] }\,\lambda$\rlap{6087}
 &     11\S
 & $<$11\S
 & $<$6\S
 &   9.7\S
 &     19\S
 & 
 & $<$15\S
 & $<$10\S
 &  
 & 
 & 
\\
  ${\rm   [OI]    }\,\lambda$6300
 &     42\S
 &     51\S
 &     24\S
 &     26\S
 &     13\UNC\S
 &      5\S
 &     62\S
 &     31\S
 &    100\S
 &     95\S
 &     13\S
\\
  ${\rm   [NII]   }\,\lambda$6583
 &    343\S
 &    407\S
 &    223\S
 &    237\S
 &    133\S
 &    128\S
 &    425\S
 &    265\S
 &    585\S
 &    586\S
 &    159\S
\\
  ${\rm   [SII]   }\,\lambda$6716
 &     96\S
 &    130\S
 &     75\S
 &     64\S
 &     37\S
 &     45\S
 &    162\S
 &    105\S
 &    285\S
 &    342\S
 &     59\S
\\
  ${\rm   [SII]   }\,\lambda$6731
 &     96\S
 &    105\S
 &     60\S
 &     57\S
 &     32\S
 &     35\S
 &    124\S
 &     83\S
 &    210\S
 &    238\S
 &     51\S
\\
  ${\rm  [ArIII]  }\,\lambda$\rlap{7136}
 &     22\S
 &     14\UNC\S
 &     11\S
 &     27\S
 &     22\S
 &  $<$10\S
 &     10\UNC\S
 &  $<$15\S
 & 
 & 
 &   $<$10\S
\\
%
%
  ${\rm   [SIII]  }\,\lambda$9531
 &    134\S
 &     68\S
 &     61\S
 &    155\S
 &    139\S
 &     29\S
 & \multicolumn{5}{c}{\dotfill not measured \dotfill not measured \dotfill }
\\
\SKIP{2}
 \HA\ flux$^c$
 &  4000\S
 &   590\S
 &   430\S
 &   110\S
 &    32\S
 &   820\S
 &   330\S
 &   260\S
 &     9\S
 &     9\S
 &  1500\S
\\
\SKIP{2}
 W$_\lambda$(\HA)\rlap{$^d$}
 &    23\S
 &    10\S
 &    37\S
 &    72\S
 &    30\S
 &    59\S
 &     8\S
 &    21\S
 &     5\S
 &     5\S
 &    24\S
\\
 W$_\lambda$([OIII])\rlap{$^d$}
 &    50\S
 &    20\S
 &    60\S
 &   300\S
 &   100\S
 &    $<$3\S
 &     8\S
 &    12\S
 &     7\S
 &     7\S
 &   
\\
\hline
\end{tabular}
\def\NOTA#1#2{\hbox{\vtop{\hbox{\hsize=0.015\hsize\vtop{\centerline{#1}}}}
      \vtop{\hbox{\hsize=0.97\hsize\vtop{#2}}}}}
\vskip1pt
\NOTA{$^{(1)}$}
{ Dereddened fluxes relative to \HB=100, extinctions are computed 
imposing \HA=290 and the adopted visual extinctions are given
at the top of each column. Blank entries are undetected lines 
with non-significant upper limits}
\NOTA{$^a$}{ \HB\ is weak and the derived extinction is therefore
uncertain.}
\NOTA{$^b$}{ Possibly overestimated due to contamination by foreground 
gas with lower extinction. }
\NOTA{$^c$}{ Dereddened flux, units of $10^{-15}$ erg cm$^{-2}$ s$^{-1}$}
\NOTA{$^d$}{ Equivalent widths in \AA\  of nebular emission lines. }
\end{flushleft}
\vskip-16pt
\end{table*}

\section{ Results }

\subsection{ Line fluxes and ratios }

The quasi-complete spectra of the nucleus and of knot C are shown
in Figs~\ref{spec_nuc}, \ref{spec_knc}  respectively, and
the derived line fluxes are summarized in Table~\ref{tab_obs1}. 
Dilution by a stellar continuum is particularly strong in the nuclear
spectrum where the equivalent width of [OIII]\L5007 
is only 50~\AA\  (cf. Table~\ref{tab_obs2}) and a factor
of $\sim$10 lower than found in typical Seyfert 2's.
The stellar contribution is normally estimated and subtracted
using either off--nuclear spectra extracted from the same 2D long 
slit frames, or a suitable combination of spectra of non--active
galaxies used as templates  (e.g. Ho \cite{ho96}, Ho et al. \cite{ho97}).
However, neither of the methods proved particularly useful because 
line emission contaminates the stellar emission all along the slit, 
and we could not find any template which 
accurately reproduces the prominent stellar absorption features 
typical of quite young stellar populations.
The fluxes of weak lines ($<$5\% of the continuum) in the
nucleus are therefore uncertain and, in a few cases, quite different
than those reported in \cite{O94}, the largest
discrepancy being for [NI] which is a factor of 2 fainter here.

The spectrum of knot C has a much more 
favourable line/\-continuum ratio and shows many faint lines which
are particularly useful for the modelling described in 
Sect.~\ref{photion_model}.

\subsection{Spatial distribution of emission lines}
\label{spatial_line_distribution}
 
\begin{figure}
\centerline{\resizebox{8.78cm}{!}{\includegraphics{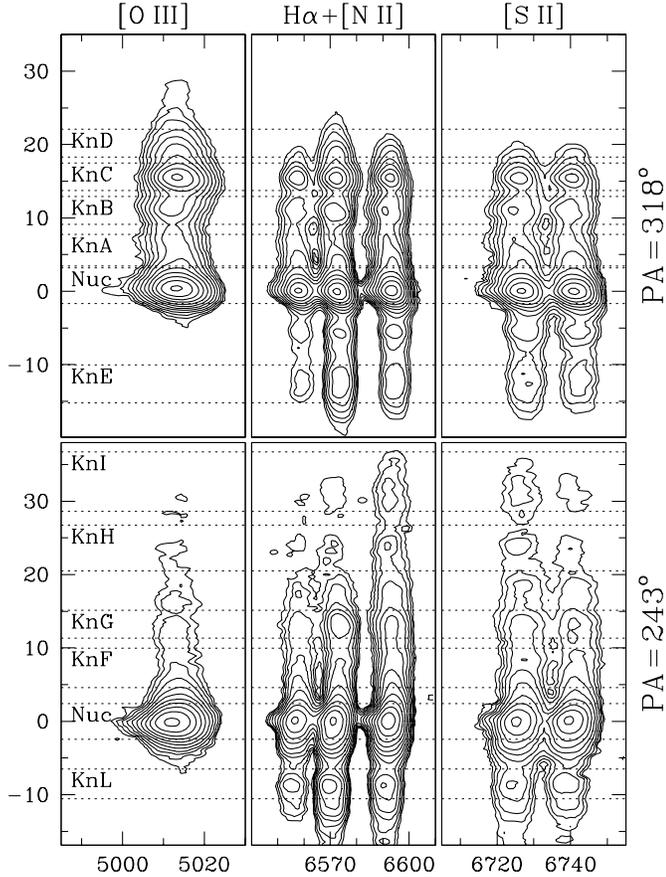}}}
\caption{
Intensity contour plots in the position--\L\ 
plane. 
The long slit spectra are continuum
subtracted and the levels are logarithmically spaced by 0.2 dex.
The ordinate are arc-sec from the
\HA\ peak along the two slit orientations (cf. Fig.~\ref{showslit}).
The dashed lines show the regions where the spectra displayed in
Figs.~\ref{spec_nuc}, \ref{spec_knc}, \ref{spec_all} were extracted.
}
\label{velcont}
\end{figure}

\begin{figure}
\centerline{\resizebox{\hsize}{!}{\includegraphics{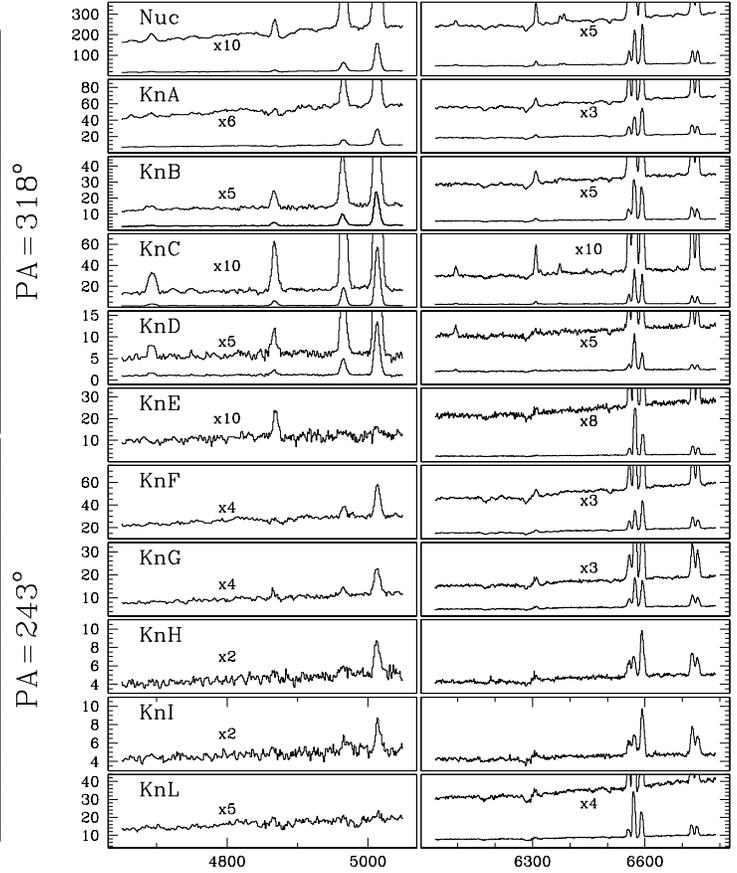}}}
\caption{
Spectra of the various knots (cf. Figs.~\ref{showslit}, \ref{velcont})
at selected wavelength ranges including [OIII], \HB, HeII (left panels)  and
[FeVII], [OI], [NII], \HA, [SII] (right hand panels). 
Fluxes are in units of $10^{-16}$ erg cm$^{-2}$ s$^{-1}$ \AA$^{-1}$ and
\L's are in \AA.
The spectra are also scaled by a factor given in the plots
to show faint features.
}
\label{spec_all}
\end{figure}

\begin{figure}
\centerline{\resizebox{8.8cm}{!}{\includegraphics{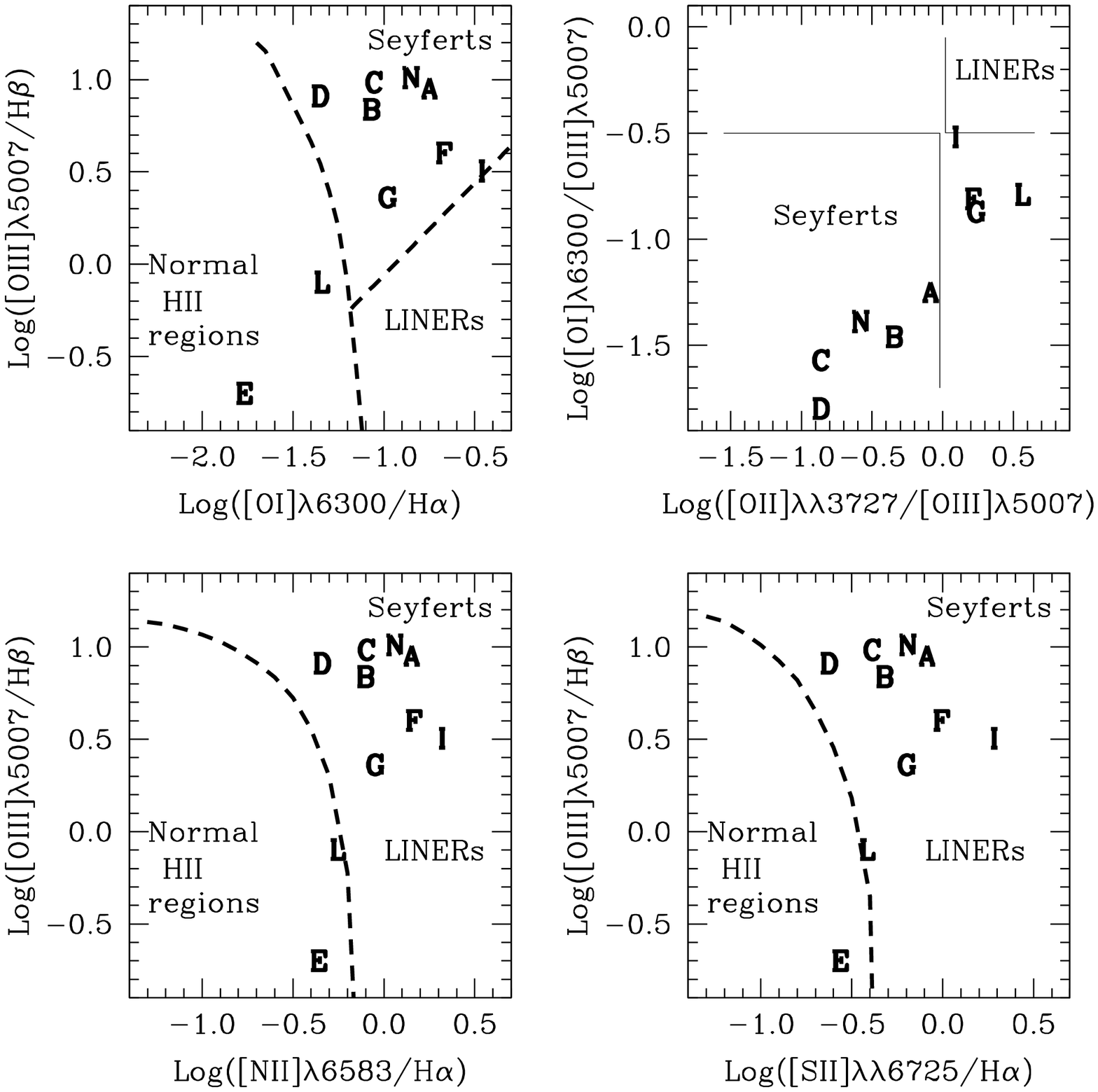}}}
\caption{
Spectral classification diagrams (from 
Veilleux \& Osterbrock \cite{veilleux87} and Baldwin et al. \cite{baldwin81})
using the line ratios observed in the various knots.
Note that ``N'' refers to the nuclear spectrum while ``I'' is the average
of the H and I knots which have similar line ratios (cf. Table 2).
}
\label{knotdiag}
\end{figure}

The spatial variation of the most important lines is visualized
in Figs.~\ref{velcont}, \ref{spec_all}
  which show contour plots of the continuum
subtracted long slit spectra and selected spectral sections of the
various knots respectively. 
The fluxes are summarized in Table~\ref{tab_obs2} together
with the extinctions which were derived from hydrogen recombination lines
assuming standard case-B ratios (Hummer \& Storey \cite{hummer}).
A remarkable result is the large variations of the
typical line diagnostic ratios 
[OIII]/\HB, [OI]/\HA, [NII]/\HA\ and [SII]/\HA\  which are plotted
in Fig.~\ref{knotdiag} and range
from values typical of high excitation Seyferts 
(nucleus, knots A, B, C, D), to
low excitation LINERs (knots H, I) and normal HII regions (knots E, L).
Another interesting result is the steep extinction gradient between
the regions outside (knots C, D, H, I) and those close to the
galactic disk (nucleus and knots A, E, L). However,
a comparison between the \BG\ map 
(Moorwood \& Oliva \cite{invited94}), the \HA\ images 
(\cite{M94}) and the observed Br$\alpha$ flux from the whole
galaxy (\cite{M96}), do not show evidence
 of more obscured ($\AV\!\sim\!10\!-\!30$) ionized regions
such as those observed in NGC4945 and other starburst galaxies
(e.g. Moorwood \& Oliva \cite{moorwood88}). Nevertheless, 
these data cannot exclude the presence of deeply embedded ionized
gas which is obscured even at 4$\mu$m (i.e. \AV$>$50 mag).

Particularly interesting is the variation of the line
ratios between the adjacent knots C and D.
The ratio [FeVII]/\HB\  is a factor of
2 larger in knot D than in C, but this most probably reflects
variations of the iron gas phase abundance (see also Sect.~\ref{iron}).
Much more puzzling is the spatial variation of
the low excitation lines [OI], [SII], [NII] which
drop by a factor 1.8, while the high excitation lines HeII, [OIII],
[NeIII], [ArIII] together with the [SII] density sensitive ratio and
[OIII]/[OII] vary by much smaller amounts
(cf. Table~\ref{tab_obs2}). This cannot therefore be explained
by variations of the ionization parameters, which should first of all
affect the [OIII]/[OII] ratio.
A possible explanation for this is discussed in Sect.~\ref{model_others}.

\subsection{Diagnostic line ratios}
\label{diagnostic}

Temperature and density sensitive line ratios are summarized in 
Table~\ref{tab_diagnostic}.
Note that only a few values are available on the nucleus because of
the strong stellar continuum which prevents the measurement of 
faint lines. 

\subsubsection{ Electron temperature }

The relatively large \Te(OIII), slightly lower \Te(SIII) and  much cooler
[NII], [SII] temperatures are typical of gas photoionized 
by a ``typical AGN'', i.e. a spectrum characterized by 
a power law continuum with a super--imposed 
UV bump peaked at $\approx$50--100 eV 
(e.g. Mathews \& Ferland \cite{mathews87}).
As \Te(OIII) mainly depends on the
average energy of $h\nu\!<\!54$ eV ionizing photons, ``bumpy'' spectra,
which are quite flat between 13 and 54 eV, yield hot [OIII].
The lower ionization species, such as [NII] and [SII], mostly form in
the partially ionized region,
heated by soft X--rays, whose temperature
cannot exceed 10$^4$ K due to the powerful cooling by collisionally
excited Ly$\alpha$ and 2--photon emission.
The contrast between the OIII and NII temperatures can be further
increased if sub--solar metallicities are adopted, because [OIII]
is a major coolant while [NII] only plays a secondary role in
the cooling of the partially ionized region.

An alternative explanation for the OIII, NII etc. temperature differences 
is to assume that part of the line emission arises from density bounded
clouds. In this case there is no need to adopt a ``bumpy'' AGN spectrum
and detailed models, assuming a pure power law ionizing continuum,
were developed by Binette et al. (1996, hereafter \cite{B96}).
However, it should be
kept in mind that \Te(OIII)$>$\-\Te(NII) does not  necessarily
indicate the presence of density bounded clouds.

\subsubsection{ Electron density}

There is a clear trend between \Ne\  and excitation of the species
used to determine the density.
 
The [SII] red doublet yields densities lower than [OII] which
is compatible with a single-density cloud for the following reason.
If the flux of soft X-rays (200-500 eV)
is strong enough, then [SII] lines are mostly produced in the X-ray
heated region where the average hydrogen ionization fraction is quite
low ($\la$0.1). The lines of
[OII] on the contrary can only be produced in the transition region,
where the ionization degree is close to unity, because of the very
rapid O--H charge exchange reactions.
Hence \Ne(SII)$<$\Ne(OII) most probably indicates the presence of a strong
soft-X flux, as one indeed expects to be the case for an AGN spectrum.
This is also confirmed by the detailed modelling described below.

The higher densities in the fully ionized region are new
results because the blue [ArIV] doublet is usually too weak in AGN
spectra and the FIR [NeV] lines are only  
accessible with the ISO-SWS spectrometer (\cite{M96}).
The [ArIV] density of knot C is equal, within the errors, to that derived
by [OII] thus indicating that no large variations of densities are 
present in this cloud.

\begin{table}
\def\SKIP#1{\noalign{\vskip#1pt}}
\def\L{\hbox{$\lambda$}}
\def\LL{\hbox{$\lambda\lambda$}}
\caption{Temperature and density sensitive line ratios.}
\label{tab_diagnostic}
\begin{flushleft}
\begin{tabular}{lcc}
\hline\hline
\SKIP{1}
\ \ \ \ \  Ratio   &  Value        & Diagnostic \\
\SKIP{1} \hline \SKIP{3}
\multicolumn{3}{c}{\dotfill\  \ Nucleus\   \ \dotfill } \\
\SKIP{3}
\ [OII]\ \LL3727/\LL7325 &       25$\pm$6       & 300$<$\Ne$<$6000  \\
\ [SII]\ \L6731/\L6716 &       1.00$\pm$0.07   & 400$<$\Ne$<$800 \\
\ [NeV]$^a$\ 14.3$\mu$m/24.3$\mu$m\ \ \ \ \ \ \  &
                            1.6$\pm$0.4   & 2000$<$\Ne$<$12000 \\
\SKIP{2}
\ [SIII]\ \L9531/\L6312&       $\ga$25          & \Te$\la$14000 \\
\SKIP{7}
\multicolumn{3}{c}{\dotfill\  \ Knot C\   \ \dotfill } \\
\SKIP{3}
\ [OII]\ \LL3727/\LL7325 &       14$\pm$4    & 1400$<$\Ne$<$11000   \\
\ [SII]\ \L6731/\L6716 &       0.89$\pm$0.06   & 200$<$\Ne$<$500 \\
\ [ArIV]\ \L4711/\L4740 &      1.0$\pm$0.2        & 1500$<$\Ne$<$10000 \\
\SKIP{1}
\ [NII]\ \L6583/\L5755 &       50$\pm$13       & 9000$<$\Te$<$12500 \\
\ [OIII]\ \L5007/\L4363 &      46$\pm$9       & 14000$<$\Te$<$18000 \\
\ [SII]\ \L6731/\LL4073 &       $\la$8        & \Te$\la$12000 \\
\ [SIII]\ \L9531/\L6312&       26$\pm$7        & 11500$<$\Te$<$16000 \\
\SKIP{1} \hline 
\end{tabular}
\vskip3pt
$^a$ [NeV] FIR lines from M96
\end{flushleft}
\end{table}

\section{ Photoionization models }
\label{photion_model}

Several photoionization
models for the Circinus galaxy have been discussed in the literature.
These used line intensities and ratios measured from the central
regions of the galaxy and were
mostly directed to constraining the intrinsic shape of AGN
photoionizing continuum. Quite different conclusions were drawn
by \cite{M96}, who found evidence for a prominent UV bump centered
at $\simeq$70 eV, and Binette et al. (1997, hereafter \cite{B97})
who showed that equally good
(or bad) results could be obtained using a combination of radiation
and density bounded clouds photoionized by a pure power--law spectrum.
In all cases the metal abundances were assumed to be solar and no
attempt was made to constrain metallicities.

Here we mainly concentrate on knot C, an extranuclear
cloud whose rich emission line spectrum and simple geometry 
(a plane parallel region) are better suited for 
deriving physical parameters from photoionization modelling.
We first analyze the observational evidence against shock excitation
and then describe in some detail the new modelling procedure
which is primarily aimed at determining the gas metal abundances,
(cf. Sects.~\ref{details_photion} and \ref{knc_abund}).
We also analyze the problems to reproduce the observed 
[FeVII]/[FeII] and [NII]/[NI] ratios (Sects.~\ref{iron}, \ref{NII_NI}),
discuss the role of dust and reanalyze several crucial aspects of the
line emission from the brightest regions closest to the nucleus.

\subsection{ Arguments against shock excitation }
\label{against_shock}

Strong evidence against shock excitation comes from the very
low strength of [FeII] which indicates that iron is
strongly underabundant, i.e. hidden in grains,
 within the partially ionized region
(cf. Figs.~\ref{figZ}, \ref{figZnoch}).
Shocks are very effective in destroying dust grains,
and velocities as low as 100 km/s are enough to return most of 
the iron to the
gas phase (e.g. Draine \& McKee \cite{drainemckee}).
This is confirmed by observations of
nearby Herbig--Haro objects (e.g. Beck-Winchatz et al. \cite{beck94})
and supernova remnants whose near IR spectra
are dominated by prominent [FeII] IR lines
(e.g. Oliva et al. \cite{oliva89}).
Although the low Fe$^+$ gas phase abundance could,
in principle, be compatible with a
very slow shock ($\la$50 km/s), this 
falls short by orders of magnitude in producing
high excitation species.

Another argument 
comes from the HeII/\HB\  and HeII/HeI ratios
which are a factor $>$2 larger than those predicted by shocks models with
velocities $v\!\le\!500$ km/s
(Dopita \& Sutherland \cite{dopita95}).
It should be noted that,
although stronger HeII could probably be obtained by increasing
the shock speed to $\sim$1000 km/s, these velocities are 
incompatible with the observed line profiles.
More generally, the
observed line widths ($<$150 km/s, \cite{O94})
are difficult to reconcile with the fact that
only shocks faster than 300 km/s can produce prominent
high excitation lines such as [OIII]
(Dopita \& Sutherland \cite{dopita96}).
This argument becomes even stronger when interpreting the nuclear spectra
where the highest excitation lines, [FeX,XI], 
do not show any evidence of large scale motions at velocities $>$200 km/s.
(cf. \cite{O94} and work in preparation).

Finally, it is worth mentioning that detailed shock~+~photoionization
composite models recently developed by
Contini et al. (\cite{contini1998})
suggest that
shock excited gas does not contribute significantly
to any of the optical/IR lines from Circinus.

\subsection{ Details of the photoionization models }
\label{details_photion}

\subsubsection{ Single component, radiation bounded clouds }

We constructed a grid of models covering 
a wide range of 
ionization parameters ($-3.5\!<\!\log\, U\!<\!-1.5$), 
densities ($2.0\!<\!\log\, n\!<\!5.0$),
metallicities ($0.08\!<\!{\rm He/H}\!<\!0.16$, 
$-1.5\!<\!\log\, {\rm Z}/\ZSUN\!<\!0$) and
shape of the ionizing continuum which we parameterized as a combination
of a power law extending from 1 eV to 10 keV
 with index $2<\alpha<0.4$ and a black--body\footnote{
The choice of a black--body does not have any direct physical
implications. We used it just because it provides a simple way to 
parameterize the UV bump which seems to be a typical features of
AGNs (e.g. Mathews \& Ferland \cite{mathews87}, Laor \cite{laor90}) and is 
predicted by accretion disk models 
(e.g. Laor \& Netzer \cite{laor_netzer89}) }
with $5\!<\!\log T_{BB}\!<\!6$, the relative fraction of the two components
being another free parameter. All the parameters were varied randomly
and about 27,000 models were produced using 
Cloudy (Ferland \cite{cloudy}) which we slightly modified 
to include the possibility of varying the N--H charge exchange rate.
The line intensities were also computed using the temperature
and ionization structure from Cloudy and an in--house data base
of atomic parameters. The agreement with the Cloudy output
was good for all the ``standard'' species while large discrepancies
were only found for coronal species (e.g. [FeX]) whose collision strengths 
are still very uncertain and much debated.

Out of this large grid we selected about 400 models whose 
[OIII]/\HB, [OIII]/[OII]/[OI], [ArV]/[ArIV]/[ArIII],  
[SIII]/[SII], 
[OIII]\L5007/\L4363, [OII]\LL3727/\LL7325,
[SII]\L6731/\L6716, [ArIV]\L4711/\L4740
line ratios
were reasonably close to those observed in knot C. These were used
as the starting point for computing the ``good models'', with adjusted
values of relative metal abundances, which minimized
the differences between predicted and observed line ratios.
Note that to reproduce both [FeVII] and [FeII] we were
in all cases forced to vary the iron gas phase abundance between the
fully and partially ionized regions (cf. also Sect.~\ref{iron}).
The results of the best model are summarized in Table~\ref{tab_modelKNC}
and discussed in Sect.~\ref{detail_knc_model}. Note the large
discrepancy for [NI] which is overpredicted by a factor of $\sim$5,
this problem is discussed in Sect.~\ref{NII_NI}.

The most important results are the abundance
histograms shown in Figs.~\ref{figZ}, \ref{figZnoch}
 which were constructed by including all
the ``good'' models with $\chi^2_{red}\!<\!5$. Although
the choice of the cutoff is arbitrary, it should be 
noticed that variations of this parameter do not alter
the mean values, but only influence the shape of the distributions
whose widths roughly double if the $\chi^2_{red}$ cutoff is increased to
values as large as 30.

\subsubsection{ Multi-density components, radiation bounded clouds }
\label{details_multidens}

The large grid described above was also organized to have
at least 4 models with the same photon flux, continuum shape and abundances
but different densities, and these were used as starting points
to construct multi-density models. We simulated
2--density clouds by coupling all available models
with different weights and selected about 300 of them,
which were used as starting points to compute the ``good models''
following the same procedure adopted for the single-density case.
We also simulated clouds with $\ge$3 density components,
but these results are not 
included here because this complication had virtually no effect 
on the quality of the fit.

The most important result is that single and multi-density
models are equally good (or bad) in reproducing the observed line ratios.
Obviously, this does not necessarily imply that knot C is a single-density
cloud, but rather indicates that the stratifications
which probably exist have little effect on the available density sensitive 
line ratios. 

\subsubsection{ Mixed models with density and radiation bounded clouds }

We constructed a grid of about 10,000
models photoionized by a power law AGN continuum with
index $2\!<\!\alpha\!<\!0.3$, which turns into
$\nu^{-0.7}$ beyond 500 eV, and covering the same range of physical
parameters as the radiation bounded clouds described above.
The column density of the radiation bounded component was always large
enough to reach a temperature $<$3000 K at its outer edge.
For each model we also computed line intensities from a density
bounded cloud which we defined as a layer with thickness $\Delta\,R$
equal to 5\% of the nominal radius of the HII Str\"omgren sphere,
numerically ($f$= filling factor)
$$ \Delta\,R\simeq 2500\; {U\over \NH \, f} \ \ \ \ \ \ \ \ {\rm pc} $$
The relative contribution of density and radiation bounded regions
was set by imposing HeII\L4686/\HB=0.6
 
This approach is somewhat similar to that adopted by \cite{B96} apart from
the following details.
The radiation bounded component here is formed by high density gas,
a condition required 
to match the observed [ArIV]\L4711/\L4740, and is concentrated
within the projected size of knot C, i.e. $<$80 pc.
In practice, the two components have the same densities and see un--filtered
radiation from the AGN.

The ``good models'' were optimized, and the best abundances derived
using the the same procedure depicted above.
Worth mentioning is that most of the ``good models'' have spectral slopes
in the range 1.3--1.5 
which are in good agreement with those used by \cite{B96}.
The most important result is that these mixed models 
give similar, though slightly poorer 
(cf. Sect.~\ref{detail_knc_model}) results than
radiation bounded clouds.

\subsubsection{ The role of dust in extranuclear knots}
\label{role_of_dust}

Dust can modify 
the ionization and temperature structure because it competes
with gas in absorbing the UV photons, and because it hides
refractory elements such as iron.
Given the low ionization parameters, however,
the first effect is negligible, i.e.
the ionization structure of the fully ionized region of knot C
is not affected by dust although this
plays an important role in 
modifying the heating--cooling balance of the partially ionized region 
heated by X--rays. 

Cooling: the refractory species hidden in the grains cannot contribute
to the line cooling and this effect is correctly computed by Cloudy.
For example, depleting Fe on grains produces higher
gas temperature and stronger [OI], [SII] etc. lines,
because it suppresses the near IR lines of [FeII]
which are among the major coolants for quasi--solar Fe/O
gas phase abundances. 

Heating: the metals hidden in the dust still contribute 
to the heating of the gas because the
X--rays have energies much larger than the binding energy
of the grains, and cannot therefore recognize
metals in dust from those in the gas phase. 
 
The major problem is that Cloudy (and probably other photoionization 
models) does not include the X--ray heating from the grain metals
and therefore underestimates the temperature of the partially ionized
region and hence may predict weaker fluxes of [OI], [SII] and other
low excitation lines.
Therefore, the models for knot C, having most of iron
depleted on grains, are not fully self--consistent 
and most probably require a too high flux of X--rays to reproduce e.g.
[OI]/[OIII]. This may imply that the ``true'' models 
should have somewhat softer spectra than those of Fig.~\ref{agn_cont}.

\subsection{ Best photoionization models of knot C:
the shape of the AGN continuum and the role of density bounded clouds}
\label{detail_knc_model}

Table~\ref{tab_modelKNC}
lists the results of the best photoionization models, selected 
from the several hundred which provided a reasonable fit
to the spectrum of knot C. Note that the results of multi-density clouds
(Sect.~\ref{details_multidens}) are not included
because they are virtually indistinguishable from those of single-density
models. 

\begin{figure}
\centerline{\resizebox{8.2cm}{!}{\includegraphics{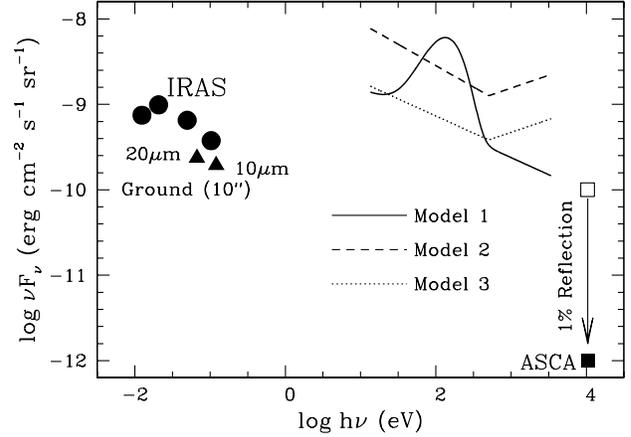}}}
\caption{
The AGN ionizing continua
used to compute the best models 
(Table~\ref{tab_modelKNC}) are plotted together with
the observed X and IR continua.
The solid line refers to a single density, radiation bounded component
which provides a good fit for all lines but [NI].
The broken curves are for combinations of density and 
radiation bounded clouds, but these models cannot
simultaneously
reproduce the [OII]/[OIII] and [ArV]/[ArIII] ratios.
 The dashed curve provides the best fit for
high excitations lines while the dotted curve best reproduces the 
OI/OII/OIII ionization balance 
(see Sect.~\ref{detail_knc_model} for details).
Note that the curves show the spectra seen by knot C, and should
be scaled by a coefficient which takes into account the ``intrinisic beaming''
of the AGN ionizing radiation (a factor of 2 for an optically thick disk)
before being compared with the observed points. See Sect.~\ref{agn_budget}
for a discussion of the AGN energy budget.
}
\label{agn_cont}
\end{figure}

Fig.~\ref{agn_cont} shows the AGN continuum adopted for the
``best models'' of Table~\ref{tab_modelKNC}.
Radiation bounded models require a prominent UV bump
centered (in $F_\nu$) at about 100 eV.
This is slightly bluer than that found by
\cite{M96} and somewhat harder than model spectra
of more luminous accretion disks (cf. e.g.
Laor \cite{laor90}), but in qualitative agreement with the predicted
dependence of spectral shape with AGN luminosity, i.e. that
lower luminosity accretion disks should have harder spectra
(Netzer et al. \cite{netzer92}).

\begin{table}
\caption{Photoionization models for knot C}
\label{tab_modelKNC}
\def\SKIP#1{\noalign{\vskip#1pt} }
\def\MYBOX#1#2{\hbox to 100pt{#1 \hfil #2}}
\def\emark{\rlap{(!)} }
\def\EMARK{\rlap{(!!)} }
\begin{flushleft}
\begin{tabular}{lccc}
\hline\hline\SKIP{1}
 Adopted parameters & Model 1  & Model 2 & Model 3\\
\SKIP{4} \hline \SKIP{2}
\hglue 10pt Element & \multicolumn{3}{c}{Abundances$^{(1)}$}  \\
\SKIP{1}
 Helium           & 11.1  & 11.1 & 11.1 \\
 Nitrogen         & 8.05  & 8.20 & 8.05 \\
 Oxygen         & 8.17 & 8.27 & 8.17 \\
 Neon           & 7.27  & 7.37 & 7.27 \\
 Sulphur        & 6.98  & 6.90 & 6.75 \\
 Argon          & 6.38  & 6.45 & 6.20 \\
 Silicon        & 6.85 & 6.95 & 6.85 \\
 Fe low-ion$^a$  & 5.80 & 5.80 & 5.80 \\
 Fe high-ion$^b$  & 6.90 & 7.00 & 7.00 \\
\SKIP{04}  \hline \SKIP{2}
  Gas density ($n_{\rm H}$) & 2000 & 3000 & 2000 \\
  Ionizing continuum$^c$ 
           & UV-bump & $\nu^{-1.5}$ & $\nu^{-1.4}$ \\
  Ionization parameter$^d$ ($U$) 
           & 3.5 $10^{-3}$  & 4.7 $10^{-3}$ & 2.0 $10^{-3}$ \\
  Fraction radiation bounded$^e$ 
           & 100\% &  4\% & 3\% \\
  $F$(\HB)/$F_{\rm RB}$(\HB)$^f$ 
           & 100\% &  6\% & 6\% \\
\SKIP{04} \hline\SKIP{2}
\MYBOX{Line ratio$^{(2)}$}{Observed}
   & Model 1 & Model 2 & Model 3\\ 
\SKIP{1}\hline\SKIP{3}
\MYBOX{HeII/\HB}{ 0.60$\pm$0.08 } &  0.60 & 0.60 & 0.60 \\
\MYBOX{HeII/HeI}{ 6.2$\pm$0.9 } & 6.8 & 5.8 & 6.3 \\
\SKIP{3}
\MYBOX{[NII]/\HA}{ 0.80$\pm$0.08 }  & 0.83 & 0.79 & 0.77 \\
\SKIP{0}
\MYBOX{[NII]/[NI]}{ 34$\pm$7 } & 8.0\EMARK & 6.5\EMARK & 4.7\EMARK \\
\MYBOX{\L6583/\L5755}{ 50$\pm$13} & 41 & 57 & 39 \\
\SKIP{3}
\MYBOX{[OIII]/\HB}{9.7$\pm$1} & 10 & 9.8 & 9.3 \\
\MYBOX{\L5007/\L4363}{46$\pm$9} & 31\emark & 43 & 42 \\
\MYBOX{[OII]/[OIII]}{0.14$\pm$0.02 } & 0.12 & .057\EMARK & 0.12 \\
\MYBOX{[OI]/[OIII]}{.027$\pm$.003}  & .027 & .024 & .028 \\
\MYBOX{\LL3727/\LL7325}{14$\pm$4}  & 14 & 12 & 13 \\
\SKIP{3}
\MYBOX{[NeIII]/\HB}{0.66$\pm$0.1}   & 0.68 & 0.70 & 0.73 \\
\SKIP{3}
\MYBOX{[SiVI]/\BG}{ $<$5} &  2.2 & 1.5 & 0.12 \\
\SKIP{3}
\MYBOX{[SII]/\HA}{ 0.19$\pm$0.02} & 0.20 & 0.20 & 0.19 \\
\MYBOX{\L6731/\L6716 }{ 0.89$\pm$0.06} & 0.95 & 0.88 & 0.91 \\
\MYBOX{[SIII]/[SII] }{ 2.8$\pm$0.4} & 3.0 & 2.4 & 2.6 \\
\MYBOX{\L9531/\L6312 }{ 26$\pm$7} & 20 & 22 & 21 \\
\SKIP{3}
\MYBOX{[ArIII]/\HA }{ .092$\pm$.011}   &  .089 & .098 & .092 \\
\MYBOX{[ArIV]/[ArIII] }{ 0.37$\pm$0.1} & 0.56 & 0.57 & 0.25 \\
\MYBOX{[ArV]/[ArIII] }{ 0.24$\pm$0.03} & 0.17 & 0.17 & .032\EMARK \\
\MYBOX{\L4711/\L4740 }{ 1.0$\pm$0.2} & 0.88 & 0.96 & 0.88 \\
\SKIP{3}
\MYBOX{[FeVII]/\HA }{ .033$\pm$.005} & .031 & .030 & .005\EMARK  \\
\MYBOX{[FeVI]/[FeVII] }{ $<$0.7} & 0.70 & 1.2 & 4.1\EMARK \\
\MYBOX{[FeX]/[FeVII] }{ $<$0.2} & .045 & .030 & .005  \\
\MYBOX{[FeII]/\BG }{ $\approx$1.5} & 1.6 & 1.8 & 1.8 \\
\SKIP{2} \hline
\end{tabular}
\def\NOTA#1#2{\hbox{\vtop{\hbox{\hsize=0.030\hsize\vtop{\centerline{#1}}}}
      \vtop{\hbox{\hsize=0.97\hsize\vtop{#2}}}}}
\NOTA{ $^{(1)}$ }{ 12+log(X/H) where X/H is the absolute abundance by number }
\NOTA{ $^{(2)}$ }{ Abbreviation used: HeII=\L4686, HeI=\L5876,
[NII]=\L6583, [NI]=\LL5200, [OIII]=\L5007,
[OII]=\LL3727, [OI]=\L6300, [NeIII]=\L3869, [NeV]=\L3426,
[SiVI]=\L19629, [SiVII]=\L24827,
[SII]=\L6731, [SIII]=\L9531, [ArIII]=\L7136,
[ArIV]=\L4740, [ArV]=\L7006, [FeVII]=\L6087,
[FeVI]=\L5146, [FeII]=\L16435}
\NOTA{ $^a$}{ Gas phase abundance in the partially ionized region }
\NOTA{ $^b$}{ Gas phase abundance in the fully ionized region }
\NOTA{ $^c$}{ UV--bump has $T_{BB}\!=\!4\,10^5$ K and $\alpha$=1.4, 
cf. Fig.~\ref{agn_cont} }
\NOTA{ $^d$}{ $U=Q(H)/4\pi R^2 n_{\rm H} c$}
\NOTA{ $^e$}{ Models 2,3 include density bounded clouds with the
same density and $U$ as radiation bounded regions, cf. Sect. 4.6.2 }
\NOTA{ $^f$}{ Flux relative to that produced by a radiation bounded model }
\end{flushleft}
\end{table}
The  amplitude of the UV bump
could be significantly decreased by relaxing the assumption
on the gas density distribution and, with a properly tuned combination
of density and radiation bounded clouds, one may probably get
a similarly good fit with a power law.
It is nevertheless instructive to analyze why our mixed models provide
a worse fit to the data and, in particular,
are unable to simultaneously reproduce the OII/OIII balance and
the [ArV]/[ArIII] and [FeVII]/[FeVI] ratios
(cf. last two columns of Table~\ref{tab_modelKNC}).
Model \#2 has a high ionization parameter and correctly
predicts high excitation lines but underestimates all [OII] lines while
model \#3, with a lower value of U, correctly reproduces the
[OII]/[OIII] ratio but predicts very faint high excitation lines.
The main reason for the above difference is that, for a given gas density,
the [OII]/[OIII] ratio is a measure of the flux of soft ionizing
photons ($h\nu$=13--54 eV), while [ArV]/[ArIII] and [FeVII]/[FeVI]
depend on the flux of harder radiation ($h\nu\!\ga\!6$ 80 eV).
The spectrum of knot C is characterized by a quite large [OII]/[OIII] ratio,
which points towards low fluxes of soft photons, and strong high
excitation lines, which require large fluxes of hard photons. These
somewhat contradictory
requirements can be easily satisfied by a spectrum which steeply rises
beyond the Lyman edge, and peaks at $\approx$100 eV, i.e. a spectrum
similar to that of Model \#1.  \\

Independently on the detailed results of the models, the following
arguments indicate that density bounded clouds may indeed play
an important role.\\
-- Ionizing spectra with pronounced
UV--bumps tend to produce too high [OIII] temperatures and
model \#1  is indeed quite hot, though still compatible (within 1.5$\sigma$)
with the somewhat noisy measurement of [OIII]\L4363
(cf. Fig.~\ref{spec_knc} and Tab.~\ref{tab_modelKNC}).
Curiously, though, the result of Model \#1  is the opposite of that obtained
by most previous models in the literature which failed to
produce hot enough [OIII].
Therefore, the classical ``[OIII] temperature  problem'' may just reflect
the fact that past models were mostly biased toward 
high metallicities and rather flat (i.e. without strong bumps) continua.  \\
-- The \HB\  flux from knot C is only $\simeq$5\%  of that expected
if the gas absorbed all the ionizing photon impinging on
the 2\arcsec~x~2\arcsec (40~x~40 pc$^2$) geometrical cross--section
of the cloud.
Such a small ``effective covering factor''
could, in principle, be obtained by assuming a suitable distribution of
radiation bounded, geometrically thin clouds or filaments. However,
density bounded clouds seem to provide a more self--consistent interpretation
because the measured value is remarkably close to the 6\% predicted by models
\#2 and \#3  (cf. note $f$ of Tab.~\ref{tab_modelKNC}).\\
-- The variation of line ratios between the different knots cannot be
explained by radiation bounded clouds illuminated by the same ionizing
continuum, but require e.g. intrinsic beaming of the AGN continuum and/or
filtering by matter bounded clouds closer to the nucleus.
Mixed models could give a more natural
explanation to the spatial variation of line ratios (see also \cite{B96}).

\subsection{Energy budget of the AGN }
\label{agn_budget}

\begin{table}
\def\SKIP#1{\noalign{\vskip#1pt}}
\def\MYBOX#1#2{\hbox to 155pt{#1 \hfil (#2)}}
\caption{Global properties of the AGN ionizing continuum$^{(1)}$}
\label{tab_globalAGN}
\begin{flushleft}
\begin{tabular}{lr}
\hline\hline
\SKIP{1}
\MYBOX{Ionizing flux seen by knot C$^a$}{cm$^{-2}$\ s$^{-1}$} & 
                   $(1\!-\!4) \times 10^{11}$ \\
Effective covering factor of knot C$^b$ & $\sim$5\%   \\
\MYBOX{$Q$(H)$_{\rm AGN}^c$}{s$^{-1}$} & $(0.5\!-\!2.0) \times 10^{54}$ \\
\MYBOX{Observed recombination rate$^d$}{s$^{-1}$} & $2\cdot10^{52}$ \\
Fraction of $Q$(H)$_{\rm AGN}$ intercepted by gas$^e$ & $\sim$1\%    \\
\MYBOX{Total AGN ionizing luminosity$^f$}{\LSUN} &  $\sim2\cdot10^{10}$      \\
\MYBOX{Observed FIR luminosity (\LFIR)$^g$}{\LSUN} & $\simeq1.2\cdot10^{10}$    \\
\SKIP{1}
\hline
\end{tabular}
\vskip3pt
\def\NOTA#1#2{
\hbox{\vbox{\hbox{\hsize=0.030\hsize\vtop{\centerline{#1}}}}
      \vbox{\hbox{\hsize=0.97\hsize\vtop{\baselineskip2pt #2}}}}}
\NOTA{$^{(1)}$} {Assuming a distance of 4 Mpc}
\NOTA{$^a$}{ Required to reproduce the $U$--sensitive and density sensitive
line ratios measured in Knot C
(cf. Sects.~\ref{detail_knc_model},~\ref{agn_budget})
}
\NOTA{$^b$}{ Ratio between the covering factor and the 
$\simeq\!40\!\times\!40$ pc$^2$ geometric cross section of the knot
}
\vskip3pt
\NOTA{$^c$}{ Total number rate of AGN ionizing photons,
assuming emission from an optically thick disk
(cf. Sect.~\ref{agn_budget})
}
\NOTA{$^d$}{ From $L$(Br$\alpha$)=$2.5\,10^5$ \LSUN\ (M96)}
\NOTA{$^e$}{ Assuming that half of the observed Br$\alpha$
emission is produced by the starburst ring}
\NOTA{$^f$}{ Assuming an average photon energy of 50 eV}
\NOTA{$^g$}{ From Siebenmorgen et al. (\cite{siebenmorgen}) }
\end{flushleft}
\end{table}
The ionizing photon flux seen by
knot~C is constrained by the $U$--sensitive and density sensitive
line ratios which yield
$$ \Phi_{ion}=U\cdot\NH\cdot c \simeq\ (1-4) \times 10^{11} \ \ {\rm cm}^{-2} $$
This can be translated into $Q$(H)$_{\rm AGN}$, the total number rate
of ionizing photons from the AGN, once the angular distribution of the
UV ionizing radiation is known. Assuming that it arises from a
geometrically thin, optically thick accretion disk, one finds
$Q(\theta)\approx\cos\theta$ (e.g. Laor \& Netzer \cite{laor_netzer89})
and
$$ Q({\rm H})_{\rm AGN} = 2\pi\,R_{\rm knC}^2\; \Phi_{ion} \ \simeq
\  {(0.5-2.0) \times 10^{54}\over\cos^2 i} \ \ {\rm s}^{-1}$$
where $R_{\rm knC}$ is the projected distance of knot C i.e. 15.5\arcsec
or $d$=300/$\cos i$ pc
from the nucleus, $i$ being the inclination angle of the cone relative to
the plane of the sky. Detailed kinematical studies indicate
$|i|\!\le\!40^\circ$ (Elmouttie et al. \cite{elmouttie}).

The AGN luminosity in the ionizing continuum is therefore
$$ L_{ion} = Q({\rm H})_{\rm AGN} <\!h\nu_{ion}\!>
     \ \simeq\  {1\!-\!4 \cdot 10^{10}\over\cos^2 i} \LSUN $$
where $<\!h\nu_{ion}\!>$ is the average photon energy which is here
assumed to be $\simeq$50 eV.
The ionizing luminosity is therefore very large but 
compatible, within the errors, with
the observed FIR luminosity
$\LFIR\!\simeq\!1.2\,10^{10}$ \LSUN\
(Siebenmorgen et al. \cite{siebenmorgen}). This
implies that the AGN emits most of its energy in the ionizing continuum
or, equivalently, that the AGN intrinsic spectrum has a prominent
peak or bump in the ionizing UV, and much weaker emission at lower energies.
This is in good agreement with computed models of accretion disks
which also predict that low luminosity AGNs, such as Circinus, should
be characterized by a quite hard ionizing continuum (cf.
Laor \cite{laor90}, Netzer et al. \cite{netzer92}).

The global properties of the AGN are summarized in Table~\ref{tab_globalAGN}
which also includes a comparison between $Q({\rm H})_{\rm AGN}$
and the observed recombination
rate, based on ISO observations of the Br$\alpha$ H--recombination line
at 4.05 $\mu$m (cf. \cite{M96}). The difference is remarkably large,
with only $\simeq$1\%  of the AGN ionizing photons being accounted
for by emission from ``normal'' ionized gas. This indicates
that the bulk of the Lyman continuum radiation
either goes into ionizing regions which are obscured even at 4 $\mu$m
(i.e. \AV$>$50 mag), or is directly absorbed by grains in dusty clouds
lying very close to the AGN.

\subsection{ Iron depletion and the [FeVII]/[FeII] problem }
\label{iron}

The observed 
[FeVII]\L6087/[FeII]\L16435$\ga$1 ratio cannot be
explained using the same iron gas phase abundance in the HeIII
Str\"omgren sphere, where FeVII is formed, and in the partially
ionized region, where iron is predominantly FeII due to the
rapid charge exchange recombination reactions with neutral hydrogen.
It should be noted that this is a fundamental problem unrelated to the
details of the photoionization models and primarily reflects
the fact that the [FeVII] line has a 
very small collision strength ($\Upsilon/\omega_1\!\simeq\!0.1$)
which is factor of about 10 lower than 
the [FeII] transition.
A Fe$^{+6}$/Fe$^+\!\ga\!2$
integrated relative abundance is thus required to reproduce the observed
line ratio. This number is incompatible with the relative sizes of 
the HeIII and partially ionized
regions which are constrained by e.g. the HeII and [OI] lines.
It is also interesting to note that this problem
is even exacerbated if the shock models 
of Dopita \& Sutherland (\cite{dopita96}) are adopted
because these predict Fe$^{+6}$/Fe$^+$ ratios much smaller than
the photoionization models described above.

\begin{figure}
\centerline{\resizebox{\hsize}{!}{\includegraphics{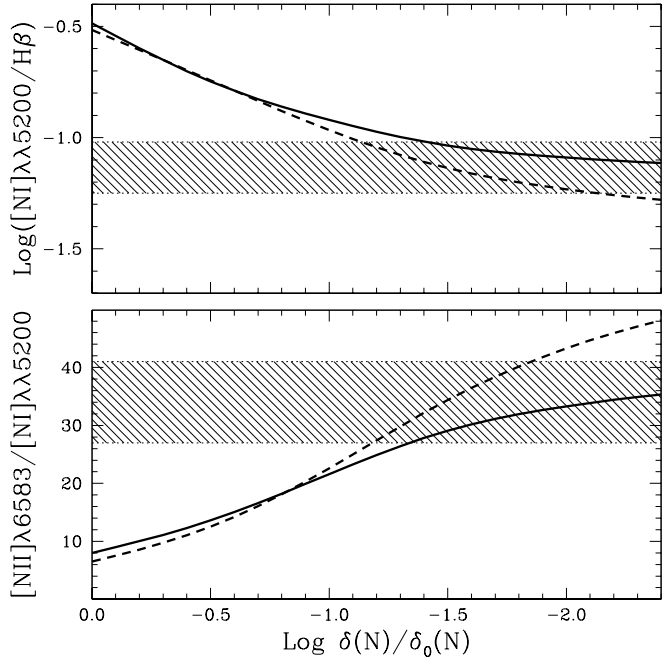}}}
\caption{
Effect of varying $\delta$(N), the rate coefficient of N--H charge exchange,
on the [NII]/[NI] and [NI]/\HB\  ratios predicted by the 
photionization models discussed
in Sect.~\ref{detail_knc_model}.
The solid line refers to model \#1 while the dashed curve
is for model \#2 (cf. Table~\ref{tab_modelKNC}), and the dashed
area show the value (with errors) measured in knot C.
The parameter $\delta_0$(N)
is the standard rate coefficient used by Cloudy.
}
\label{Nitrogen}
\end{figure}
A possible solution could be to advocate that the [FeVII] collision
strengths are underestimated by a factor of $\sim$10 which would
reconcile the observed [FeVII] and [FeII] intensities with a
low (2\% of solar) iron abundance (cf. Figs.~\ref{figZ}, \ref{figZnoch}).
However, although the collision strengths of coronal Fe lines are known
to be very uncertain, and vary by factors $>$10 depending on whether
resonances are included in the computation 
(cf. Sect. 5 of Oliva \cite{oliva_china97}),
the available collision strengths for [FeVII] 
increase by only 20\% between the old DW computations of Nussbaumer \&
Storey (\cite{Fe7-DW}) and the newer R--MAT (i.e. including resonances)
values of Keenan \& Norrington (\cite{Fe7-RMAT}).
More detailed modelling of the [FeVII] lines and spectroscopic studies
of nearby astrophysical laboratories (e.g. high excitation planetary 
nebulae) are required to clarify this issue.

The alternative possibility is to assume that the iron depletion is 
much larger in the partially than in the fully ionized region,
as already suggested by Netzer \& Laor (\cite{netzer_laor93})
and Ferguson et al. (\cite{ferguson97}). However, we
could not find any simple explanation for such a stratification
in knot C which lies far from the AGN and receives a relatively
modest flux of hard UV photons. Therefore, Fe--bearing dust cannot be 
photo--evaporated and the only mechanism to destroy grains is sputtering.
The slow shock produced by the ionization front
which propagated outward when the AGN turned on was too
slow ($\le$40 km/s) to effectively return Fe to the gas phase
(e.g. Draine \& McKee \cite{drainemckee}).
Faster shocks ($\ge$100 km/s) are a natural and efficient source of 
sputtering but cannot explain the observations because they are
expected, and observed, to emit prominent [FeII] lines from the
dust--free recombining gas. 
A possibility to overcome this problem is a combination of 
shocks and photoionization where e.g. the dust--free gas processed by the
shock is kept fully ionized by the AGN radiation. However, this
situation is short lived because, after a few thousand years, the gas piled
up behind the shock will eventually reach a
column density high enough to become radiation bounded, and 
shield the recombining gas which will therefore show up in [FeII].
%

\subsection{ The [NII]/[NI] dilemma }
\label{NII_NI}

The photoionization models of knot C systematically overestimate
by large factors ($>$6) the strength of [NI] relative to [NII]
(cf. Table~\ref{tab_modelKNC} and Fig.~\ref{figZ}). 
Although [NI]\LL5200 has a quite low critical density ($\simeq$1500
cm$^{-3}$), this error cannot be attributed to the presence
of higher density clouds because these would also effect the [SII]
density sensitive ratio. In other words, multi-density models
which correctly reproduce the high [NII]/[NI] ratio inevitably predict
too large [SII]\L6731/\L6716 ratios. Also, we can exclude observational
errors because, in knot C, the [NI] doublet 
has an equivalent width of about 2.5 \AA\  and is only marginally affected
by blending with neighbouruing stellar absorption lines.

A possible solution is to argue that the  
rate coefficients for N--H charge exchange are overestimated,
as already suggested by e.g.
Ferland \& Netzer (\cite{ferland_netzer}).
In the partially ionized region, NII is mostly neutralized via
charge exchanges with H$^0$ and adopting lower charge
exchange efficiencies yield larger [NII]/[NI] ratios.
This is evident in Fig.~\ref{Nitrogen} where this ratio is plotted
as a function of the assumed value of $\delta$(N), the charge exchange rate
coefficient. 
Assuming a N--H charge exchange a factor of $\sim$30 lower than
presently adopted yields the correct [NII]/[NI] ratio.

Noticeably, the problem we find here is exactly the opposite of what
reported by Stasinska (\cite{stasinska84}) whose models systematically
underpredicted the [NI]/[NII] ratio in objects with low [OI]/[OII].

\subsection{ Modelling the nuclear spectrum: dusty, dust--free and
diffuse components. }
\label{model_nuc}

A puzzling aspect of the line emission from regions very close to the
nucleus
is the simultaneous presence of high (e.g. [FeXI]) and low
(e.g. [SII], [OI])
ionization species. More specifically, the images of \cite{M94} clearly show
that [SII] peaks at a distance of only $\simeq$0.5\arcsec\  or 10 pc from the
nucleus (cf. Fig. 9 of \cite{M94}).
This result is incompatible with the standard idea according to which
low excitation lines are produced in clouds with low ionization parameters.
Numerically, combining
the ionizing continuum inferred from the spectrum of knot C with
 $n$=1.2$\,10^4$ cm$^{-3}$, the highest density
compatible with the FIR [NeV] doublet (Table~\ref{diagnostic}),
yields $U\!\simeq\!0.5$
at $r$=10 pc from the AGN, an ionization parameter
far too large for the production of low excitation species.
Not surprisingly therefore, all the models so far developed for the
high excitation lines (\cite{O94}, \cite{M96}, \cite{B97})
predict that [SII] should peak much farther out and have a much lower
surface brightness than that observed.

\begin{figure}
\centerline{\resizebox{8.8cm}{!}{\includegraphics{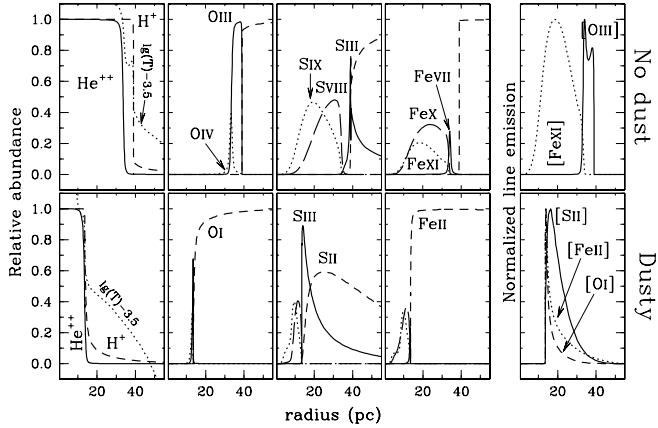}}}
\caption{
Ionization structure and spatial variation of the line emission
from two nuclear clouds exposed to the AGN
continuum of ``Model \#1''
(cf. Fig.~\ref{agn_cont})
and with identical physical parameters (\NH=$10^4$ cm$^{-3}$ and
filling factor $f$=0.025)
except for dust. The model without dust emits most of the coronal
lines, while the dusty cloud
accounts for the prominent low excitation lines observed close to the nucleus.
However, both models fail to produce enough [OIV] and
other intermediate ionization lines which therefore require a third,
more diffuse component.
See Sect.~\ref{model_nuc} for details.
}
\label{modelnuc}
\end{figure}

A simple and indeed natural solution is to assume that
the low excitation lines are formed in dusty clouds. At these large
ionization parameters dust dominates the absorption of UV ionizing
photons and, therefore, quenches the HeIII and HII Str\"om\-gren spheres.
Consequently, the X--ray dominated partially ionized region
starts at much smaller radii, and is also slightly hotter
than in the dust free model (cf. Fig~\ref{modelnuc}).
Therefore, dusty clouds have a [SII] peak much closer to the nucleus
and a $>$10 times higher surface brightness than dust--free clouds.
It should be noted, however, that the total luminosity of low excitation
optical lines is similar in the two cases, fundamentally
because the available luminosity of soft X--rays is the same.

Although the combined emission of dusty and dust--free clouds can
account for the observed emission of low ionization and
coronal species, it falls short by large factors
in producing [OIV] and other relatively
low ionization species which form within the HeIII sphere.
This is an intrinsic limitation of ``compact models'' such as those
of Fig.~\ref{modelnuc}, and can be understood as follows.
Compact models are characterized by large ``ionization parameters''
(cf. Sect. 3 of Oliva \cite{oliva_china97} for a critical
discussion on this parameter)
and therefore have large fluxes of OIV ionizing photons
which keep oxygen in higher ionization stages (mostly OVI and OVII) at all
radii inside the HeIII Str\"omgren sphere. Outside the HeIII region,
on the contrary, oxygen cannot be ionized above OIII
because most of the OIII ionizing photons have already been absorbed by HeII.
Therefore, OIV can only exist in a very narrow range of radii, just
at the edge of the HeIII sphere, and its relative abundance is therefore
very low.

In practice, we found it impossible to construct a single model which
simultaneously produces a compact coronal line region, such as that
observed in [FeXI] (\cite{O94}), and which comes anywhere close to the
[OIV]\L25.9/[OIII]\L5007$\simeq$0.3 observed ratio.
We did indeed construct many thousands of randomly
generated dusty and dust--free models, and attempted an approach similar
to that used for knot C (Sect.~\ref{details_photion})
but, in no case, could we find a model
which satisfies these contradictory constraints.
It should also be noted that \cite{B97}
independently come to a similar conclusion.

The main conclusion therefore is that, regardless of the details
of the models, the nuclear spectrum and line spatial distribution
can only be modeled by adding
a third ``diffuse'' component (e.g. with a lower filling factor)
to the dusty and dust--free clouds discussed above and depicted in
Fig.~\ref{modelnuc}.
Given the large number of free parameters, we abandoned the idea of
using photoionization models to
constrain abundances and other physical properties of the gas,
We made some attempt to verify that a mixture of clouds exposed to the same
continuum, and with the same abundances as Model \#1 of knot C
(Table~\ref{tab_modelKNC} and Fig.~\ref{agn_cont}) could
reasonably well reproduce the observed
properties of the nuclear spectrum. However, the results are not too
encouraging and, apart from the much improved [SII] surface brightness and
[OIV]/[OIII] ratio, the solutions are
not significantly better than those already discussed
by \cite{M96} and \cite{B97},
and are not therefore discussed here.

\subsection{ Modelling other extra--nuclear knots}
\label{model_others}

An analysis similar to that used for knot C was also applied to
the other extra--nuclear knots
using the more limited number of lines available in their spectra.
The results can be summarized as follows.

The abundances derived in the Seyfert-type knots A, B, D, G, F (cf.
Fig.~\ref{knotdiag}) are similar those found in knot C
but affected by much larger errors because of the more limited numbers 
of lines available for the analysis. In particular, the density
sensitive [ArIV] doublet and the U--sensitive
[ArV] line is not detected in any of these knots, and
the reddening correction for [OII]$\LL3727$ could be very uncertain 
in the high extinction regions (cf. note $b$ of Table~\ref{tab_obs2}).

We also attempted to verify if the observed line ratios in these
knots could be explained as photoionization by the same AGN continuum 
seen by knot C but could not find any satisfactory solution using
radiation bounded clouds exposed to the same
continuum. Adding matter bounded clouds alleviates
the problem (as already stressed by \cite{B96})
but requires an {\it ad hoc} choice of their photoelectric
opacities, i.e. the radius at which the ionization structure is cut.
In particular, explaining the drop of low excitation lines
between the adjacent knots C and D (cf. Sect.~\ref{spatial_line_distribution}) 
requires matter bounded clouds
cut at about 1.2$\times$ the HII Str\"omgren radius, the exact position
of the cut depending on the assumed shape of the AGN continuum.
Another parameter affecting the ratio of low--to--high 
excitation lines (e.g. [OI]/[OIII]) is the iron gas phase abundance
which influences the cooling of the partially ionized region
(cf. Sect.~\ref{role_of_dust}). 
If iron is more abundant in knot D, as indicated by its
stronger [FeVII] line emission
(cf. Table~\ref{tab_obs2} and Sect.~\ref{spatial_line_distribution}),
than [OI], [SII], [NII] could be depressed by the increased [FeII]
cooling. 

The abundances derived for the LINER--like knots (H, I) are 
very uncertain ($\pm$1.3 dex at least).
Their spectra are not compatible with illumination 
from the same continuum
seen by knot C but require a harder (i.e. more X rays relative to 
13--80 eV photons) spectrum which could be in principle obtained
by filtering the AGN continuum through an absorber with a 
carefully tuned photoelectric opacity.
Alternatively, the weak (in surface brightness) spectrum of these
low excitation knots could be explained by shock excitation, in which
case one expects [FeII]\L16435/\HB$\simeq$1 and a factor $>$10
larger than in the case of pure photoionization.

Finally, the oxygen lines in the highly reddened HII--like knots (E, L) 
are too weak to allow any reliable abundance analysis.

\begin{figure}
\centerline{\resizebox{8.8cm}{!}{\includegraphics{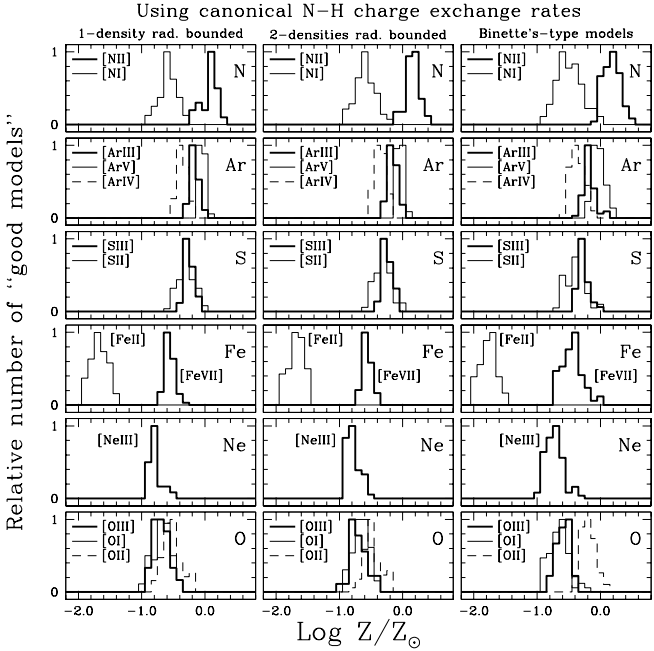}}}
\caption{
Element abundances (in log of solar units) as derived from a 
comparison between the observed line strengths in knot C, and the
predictions of a large
grid of randomly generated photoionization models. 
The three columns refer to clouds with different gas density distributions,
and the ``good models'' are
those coming closer to reproducing the observed line ratios.
See text, Sects.~\ref{knc_abund} and \ref{details_photion} for details.
}
\label{figZ}
\end{figure}
\begin{figure}
\centerline{\resizebox{8.8cm}{!}{\includegraphics{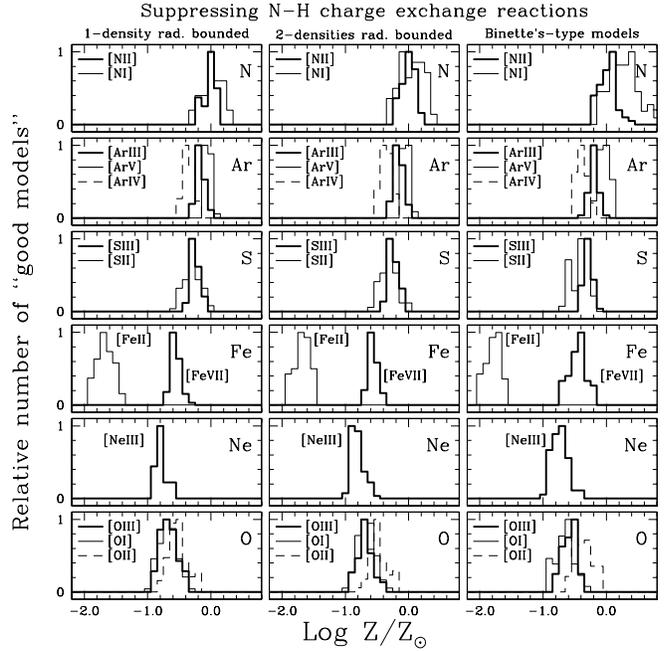}}}
\caption{
Same as Fig.~\ref{figZ} but using photoionization models where
the rate coefficient for N--H charge exchange reactions were arbitrarily
set to zero. 
Note that this  provides a much better match for the [NII]/[NI] line ratio
while has a small effect on the Nitrogen abundance derived from 
[NII] lines.
See text, Sects.~\ref{knc_abund} and \ref{details_photion} for details.
}
\label{figZnoch}
\end{figure}

\section{Discussion }

\subsection{Element abundances}
\label{knc_abund}

Deriving abundances of AGN clouds from photoionization models 
has generally been considered to be unreliable
because the shape of the AGN ionizing continuum and the
density distribution of the emitting regions (clouds) are both
basically unknown.
The more typical approach therefore has been
to assume a metallicity and use photoionization models
to constrain the AGN spectrum and/or the gas density distribution,
or just to demonstrate that the gas is photoionized.
Although explicit statements concerning the nitrogen abundance are often found
in the literature (e.g. Storchi-Bergmann \& Pastoriza \cite{storchi89},
Simpson \& Ward \cite{simpson96}) these are based on a very
limited choice of photoionization model parameters and in most cases
find oxygen abundances close to solar, in disagreement with what is
derived here. 
The only other piece of
work which covers a model parameter range comparable to that presented
here is that by Komossa \& Schulz (\cite{komossa97}) who 
analyzes a much more limited numbers of lines, e.g. do not include
[ArIV,V], in a large sample of Seyferts. They find that, on average,
oxygen is underabundant by a factor of $\sim$2 and that the N/O 
ratio is only a factor of 1.5--2.0 above the solar value.
We believe that the analysis presented here leads to more reliable
metallicity estimates.

The results presented here indicate that, in spite of the above uncertainties,
reliable metallicities can indeed be derived from spectra including a large
enough number of lines, such as that of knot C. 
Our method has been described in Sect.~\ref{details_photion}
and, in short, is based on the computation of a large number of
photoionization models with
a minimal personal bias and preconceived ideas on 
the AGN ionizing spectrum and the gas density distribution.
In particular, we also considered mixed models with combinations
of density and radiation bounded clouds (\cite{B96})
as well as models with multiple density components.
We spanned a very wide range of model parameters which were varied
randomly, and selected the relatively few (about 200) good models which
came closest to reproducing the observed line ratios.

The main results are summarized in Figs.~\ref{figZ}, \ref{figZnoch}
which show the 
distribution of element abundances required to reproduce the 
observed line ratios.
The most remarkable feature is that the metal abundances
are quite well constrained in spite of the 
very different assumptions made for the gas density
distribution and shape of the AGN continua.
In other words, models with very different abundances fail to match
the observed lines ratios in knot C regardless of the AGN spectral shape
and/or gas density distribution assumed.

Another encouraging result is that lines from different ionization
stages yield similar abundances which simply reflects the
fact that the models reproduce the observed line
ratios reasonably well. There are however remarkable exceptions, such as
the [NII]/[NI] and [FeII]/[FeVII] ratios which
are both predicted too high. Possible explanations for
these differences have been discussed above
(Sects.~\ref{iron}, ~\ref{NII_NI}). We stress here, however, that
the uncertainties on [NI] have little effects
on the derived nitrogen abundance
because N$^+$ is the most abundant ion within the partially ionized region.
Therefore, the best--fit N abundance decreases by only
a factor $\simeq$1.3 once the N$^+$/N$^0$ is increased to match
the observed [NII]/[NI] ratio (cf. Fig.\ref{figZnoch}).
Note also that the He/H abundance is only poorly constrained by the models,
and although models with He/H$>$0.1 are somewhat favoured, no firm
conclusion about He overabundance can be drawn from the data.\\

\begin{table}
\caption{Metal abundances in knot C$^{(1)}$}
\label{tab_Z}
\def\SKIP#1{\noalign{\vskip#1pt} }
\begin{flushleft}
\begin{tabular}{lccccc}
\hline\hline\SKIP{2}
 Element & 12+log(X/H)\rlap{$^{(2)}$} & 
   [X/H]\rlap{$^{(3)}$} & \hglue2pt\ &
\multicolumn{2}{c}{[X/O]\rlap{$^{(4)}$}} \\
 & & & & obs & model\rlap{$^{(5)}$} \\
\SKIP{2} \hline \SKIP{1}
 Nitrogen${^a}$  & 8.0 & $+0.0$ & & $+0.7$ & $+0.6$ \\
 Oxygen     & 8.2 & $-0.7$ & & -- & -- \\
 Neon       & 7.3 & $-0.8$ & & $-0.1$ & $-0.2$ \\
 Sulphur    & 6.9 & $-0.3$ & & $+0.4$ & \ n.i.\rlap{$^b$} \\
 Argon      & 6.4 & $-0.2$ & & $+0.5$ & \ n.i.\rlap{$^b$} \\
\SKIP{7}
 Stellar metallicity\rlap{$^c$} & -- & $-0.7$\rlap{$^d$} & &     & \\
\SKIP{2} \hline
\SKIP{2}
\end{tabular}
\def\NOTA#1#2{
\hbox{\vbox{\hbox{\hsize=0.030\hsize\vtop{\centerline{#1}}}}
      \vbox{\hbox{\hsize=0.97\hsize\vtop{\baselineskip2pt #2}}}}}
\NOTA{$^{(1)}$ }{ All values are $\pm$0.2 dex. Iron is not included
because its relative abundance is very uncertain (cf. Sect.~\ref{iron})}
\NOTA{$^{(2)}$ }{ Absolute abundance by number}
\NOTA{$^{(3)}$ }{ Log abundance by number relative to 
H=12.0, N=7.97, O=8.87, Ne=8.07, S=7.21, Ar=6.60, Fe=7.51,
the adopted set of solar abundances}
\NOTA{$^{(4)}$ }{ [X/O]=log(X/O)-log(X/O)$_\odot$ }
\NOTA{$^{(5)}$ }{ Predicted for a $\simeq\!3\,10^8$ yr old starburst 
(Fig.~4 of Matteucci \& Padovani \cite{matteucci93}), see 
Sect.~\ref{N_and_starburst} for details }
\vskip1pt
\NOTA{ $^a$}{ From Fig.~\ref{figZnoch}}
\NOTA{ $^b$}{ Element not included in the chemical evolution model}
\NOTA{ $^c$}{ Derived from CO stellar absorption features 
(cf. Sect.~\ref{stellar_abundance}) }
\NOTA{ $^d$}{ $\pm$0.3 dex }

\end{flushleft}
\end{table}
The derived abundances are summarized in Table~\ref{tab_Z}
where the most striking result is the large overabundance
of nitrogen relative to oxygen, +0.7 dex above the solar value, whose
implications are discussed below.

\subsection{ Comparison with other abundance estimates}
\label{stellar_abundance}

An independent estimate of metallicity can be derived from the measured
equivalent widths of CO stellar absorption features, using
the new metallicity scale proposed and successfully applied 
to young LMC/SMC
clusters  by Oliva \& Origlia (\cite{oliva_origlia98}). 
In short, the method is based on
the strength of the CO(6,3) band--head at 1.62 $\mu$m whose 
behaviour with metallicity is modelled using
synthetic spectra of red supergiants. 
The equivalent width of the stellar CO lines from the central
100~x~100 pc$^2$ of Circinus are reported in
Table 2 of Oliva et al. (\cite{oliva95}) and
yield an average metallicity of $-0.7\!\pm\!0.3$, 
a value remarkably close to the
oxygen abundance derived above (cf. Table~\ref{tab_Z}).

\subsection{ Nitrogen overabundance and starburst activity }
\label{N_and_starburst}

The nitrogen overabundance is of particular interest in view of its possible 
relationship with the (circum)nuclear starburst and
N--enrichment from material processed through the CNO cycle.
According to chemical evolutionary models of starburst events,
the N/O relative abundance
reaches a maximum value of [N/O]$\simeq\!+0.6$
(i.e. 4 times the solar value) at about $3\,10^8$ yr and remains
roughly constant for several $\times10^8$ years
(cf. Fig.~4 of Matteucci \& Padovani \cite{matteucci93}). 
The nitrogen overabundance mostly reflects the effect of the winds from 
He burning red supergiants whose surface 
composition is strongly N--enriched by gas 
dredged--up from the shell where hydrogen was burned through the CNO cycle.
The amount and temporal evolution of the N/O abundance depends on
model details, e.g. the shape of the IMF and the duration of the starburst,
as well as on poorly known
parameters such as the efficiency of the dredge--up
and the contribution of primary 
N production by massive stars (e.g. Matteucci \cite{matteucci86}).
It is however encouraging to find that the observed N/O abundance 
(Table~\ref{tab_Z})
is very close to that predicted at a time which is compatible with
the age of the starburst in Circinus (cf. Fig. 9 of Maiolino et al. 
\cite{maiolino}). 
It should also be noticed that the observed absolute abundances
are about an order of magnitude lower than the model predicted values,
 but this
can be readily explained if the starburst transformed only 
$\simeq\!10\%$ of the available gas into stars, in which case the
chemical enrichment was diluted by a similar factor.
This hypothesis is in good agreement with the fact that Circinus is
a very gas rich galaxy (e.g. Freeman et al. \cite{freeman}).

In short, the observed N/O overabundance is fully compatible with
what expected for a quite old (several $\times10^8$ yr) starburst.
The [NII]/\HA\  and other line ratios measured in the 
cone of Circinus are similar
to those observed in many others Sy2's and several
observational results indicate that relatively old starburst events 
are common in type 2 Seyferts (e.g. Maiolino et al. \cite{maiolino95}).
This may therefore indicate that nitrogen is typically overabundant in
Sy2's due to
enrichment by the starburst associated with the AGNs.
This tantalizing conclusion should be however 
verified by detailed 
spectroscopic studies and analysis of a sufficiently large number
of objects.

\section{Conclusions}

By modelling the rich spectrum of an extranuclear cloud in the ionization
cone of the Circinus galaxy
we have found that metal abundances are remarkably
well constrained, regardless of the assumptions made on the shape
of the ionizing continuum and gas distribution.
This new result may open a new and interesting field of research using
photoionization models to derive metallicities in AGNs 
which could in turn be related to the star formation
activity in the recent past, i.e. old nuclear starbursts.
In the case of Circinus, the large N/O overabundance found here is fully
compatible with what expected from chemical evolution models of starbursts
(Sect.~\ref{N_and_starburst}).

Much less encouraging are the results on the
AGN ionizing continuum whose shape cannot be constrained by
the observed line ratios but depends on the
assumed gas density distribution. Within the limits of the model
parameters spanned here we somewhat favour an AGN spectrum  with
a ``UV--bump'' but cannot exclude that, with a different and better
tuned combination of density and radiation bounded clouds, 
one could achieve similarly good fits with a power law AGN continuum
(Sect.~\ref{detail_knc_model}).

We also found that photoionization models cannot reproduce the observed
[FeVII]/[FeII] and [NII]/[NI] ratios and argued that these may reflect
errors in the collision strengths for [FeVII]
and rate coefficient of N--H charge exchange reactions.
It should be noted, however, that the problem [NII]/[NI] does not
significantly influence the derived nitrogen abundance.

Finally, our data strongly indicate that shocks cannot play any important
role in exciting the gas producing the observed line emission.

\begin{acknowledgements}
We thank Roberto Maiolino and Francesca Matteucci for useful
discussions, Luis Ho and Thaisa Storchi Bergman for providing
information on template stellar spectra, Gary Ferland for making
Cloudy available to the community, and an anonymous referee
for many comments and criticisms which have been fundamental for improving
the quality of the paper.
\end{acknowledgements}

\end{document}